# A general positivity-preserving algorithm for implicit high-order finite volume schemes solving the Euler and Navier-Stokes equations

Qian-Min Huang [a,b,c], Yu-Xin Ren [a*1], Qian Wang [d]

[a] Department of Engineering Mechanics, Tsinghua University, Beijing 100084, China

[b] CAEP Software Center for High Performance Numerical Simulation, Beijing 100088, China

[c] Institute of Applied Physics and Computational Mathematics, Beijing 100088, China

[d] Beijing Computational Science Research Center, Beijing 100193, China

**Abstract**

This paper presents a general positivity-preserving algorithm for implicit high-order finite volume schemes solving Euler and Navier-Stokes equations. Previous positivity-preserving algorithms are mainly based on mathematical analyses, being highly dependent on the existence of low-order positivity-preserving numerical schemes for specific governing equations. This dependency poses serious restrictions on extending these algorithms to temporally implicit schemes, since it is difficult to know if a low-order implicit scheme is positivity-preserving. The present positivity-preserving algorithm is based on an asymptotic analysis of the solutions near local vacuum minimum points. The asymptotic analysis shows that the solutions decay exponentially with time to maintain non-negative density and pressure at a local vacuum minimum point. In its neighborhood, the exponential evolution leads to a modification of the linear evolution process, which can be modelled by a direct correction of the linear residual to ensure positivity. This correction however destroys the conservation of the numerical scheme. Therefore, a second correction procedure is proposed to recover conservation. The proposed positivity-preserving algorithm is considerably less restrictive than existing algorithms. It does not rely on the existence of low-order positivity-preserving baseline schemes and the convex decomposition of volume integrals of flow quantities. It does not need to reduce the time step size for maintaining the stability either. Furthermore, it can be implemented iteratively in the implicit dual time-stepping schemes to preserve positivity of the intermediate and converged states of the sub-iterations. It is proved that the present positivity-preserving algorithm is accuracy-preserving. Numerical results demonstrate that the proposed algorithm preserve the positive density and pressure in all test cases, and considerably improve the robustness of the numerical schemes.

---

[1] Corresponding to ryx@tsinghua.edu.cn

# 1. Introduction

High-order numerical methods have shown significant advantages over second-order methods in simulating multi-scale flows such as turbulence and aerodynamic noise. They are superior to second-order methods in resolving flow structures such as boundary layers, vortices, shear layers and flow separations. Over recent decades, a variety of high-order methods have been developed [1]-[3], including the high-order finite volume (FV) [4]-[15], high-order finite element (FE) [16]-[19] and their extensions [20]-[26].

However, higher-order methods are not widely applied in practical applications, since there still are challenging issues to address, such as the complexity in scheme design and optimization, large CPU time and memory demands, deterioration of robustness and lack of effective shock capturing and efficient implicit time-stepping techniques [27]. Among these issues, the deterioration of robustness is a crucial one that poses severe restriction on the application of high-order methods in practical engineering problems. A commonly encountered robustness issue for a compressible flow solver is that it cannot preserve non-negativity of density and pressure, leading to an abnormal termination of simulation [28].

Besides density and pressure, there are many other variables that are non-negative, such as the volume/mass fraction in multiphase/multicomponent flows [29], turbulent eddy viscosity in turbulence models [30], water wave height in shallow water equations [31] and probability density function of particles in gas kinetic equation [32]. Therefore, it is very important to design positivity-preserving algorithms for high-order numerical schemes. In the present paper, a novel positivity-preserving algorithm is developed for high-order FV methods solving the compressible Euler and Navier-Stokes (NS) equations.

There have been many studies devoted to the development of positivity-preserving numerical schemes for compressible flows. Pertame and Shu [33] conducted early research on the positivity-preserving FV method for Euler equations and proposed a general framework. Based on this framework and the maximum-principle-satisfying high-order schemes [34], Zhang and Shu [28] proposed a positivity-preserving algorithm for arbitrary high-order accurate FV and discontinuous Galerkin (DG) methods solving Euler equations on rectangular meshes. This method was extended to the finite difference (FD) method [35], DG method on triangular meshes [36] and FV HWENO schemes [37]. These methods require the convex decomposition of volume integrals of flow quantities to preserve positivity. And a side effect of this decomposition is the reduction of time step size for maintaining stability. Hu et al. [38] proposed a positivity-preserving algorithm with less restrictive time step size for the high-order methods solving Euler equations using a simple flux limiter. Xiong [39] et al. developed a parameterized positivity-preserving algorithm for high-order FV and FD WENO schemes solving Euler equations, which was extended to the WENO schemes on unstructured grids by Christlieb et al. [40]. All aforementioned algorithms require the existence of positivity-preserving first-order schemes. For the Euler equations, it is not difficult to find such schemes. However, this is not true for the NS equations. In [42], Zhang proposed a positivity-preserving local Lax-Friedrichs type flux for the compressible NS equations, which satisfies the weak positivity property when using the SSP Runge-Kutta time integration. However, the time step should not be larger than $O\left(\operatorname{Re}\Delta x^2\right)$. Most of the positivity-preserving algorithms can be only applied together with explicit time integration schemes. Positivity-preserving algorithms for numerical methods using implicit time-stepping are very rare. Only recently, Parent [41] proposed a positivity-preserving dual-time-stepping implicit scheme for the FV methods solving Euler equations based on a Cauchy-Kowalevski procedure.

It is observed that existing positivity-preserving algorithms are mainly based on the application of certain positivity-preserving limiters to the solution polynomials or the numerical fluxes of the high-order schemes. In extreme cases, the limiters modify the high-order schemes to baseline positivity-preserving low-order schemes. Therefore, these algorithms highly depend on the existence of low-order positivity-preserving schemes. For governing equations involving viscous terms, complex equations of state or stiff source terms, the design of low-order positivity-preserving schemes becomes increasingly complicated. More seriously, it is extremely difficult to design baseline positivity-preserving schemes when implicit time integration is applied, since for an implicit scheme, a very large system of non-linear algebraic equations must be solved. It is generally impossible to estimate the positivity of the solution before the convergence is reached.

Implicit time-stepping schemes are much more efficient than their explicit counterparts in solving problems with very small allowable time step sizes determined by the CFL condition. Some examples of such problems include high Reynolds number flows with large aspect ratio grids in the boundary layer region, and turbulence and chemical reaction models involving stiff source terms. Therefore, it is of great interest to develop positivity-preserving algorithms for high-order numerical schemes in terms of the implicit temporal discretizations solving Euler and Navier-Stokes equations. To the best knowledge of the authors, no such algorithm has been reported.

This paper presents a general positivity-preserving algorithm for implicit high-order FV schemes solving Euler and Navier-Stokes equations. The present positivity-preserving algorithm is based on the physical consideration in terms of asymptotic analysis of the solutions of Euler and Navier-Stokes equations near the local vacuum minimum points. In [43], similar analysis has been carried out for the Spalart-Allmaras (SA) turbulence model, which showed that the solution of the SA model exhibits an exponential decay at the near zero minimum of the turbulent viscosity. This property was used to design the positivity-preserving algorithm for SA model. However, in [43], only steady-state flow is considered and the temporal discretization is the first-order backward Euler scheme. In the present paper, the asymptotic analysis of the Euler or NS equations is refined by keeping a forcing term. Without this forcing term, the solution decays exponentially. With the forcing term, the exponential evolution becomes more complicated. However, this analysis shows that the forcing term can be controlled to obtain a positivity-preserving solution. The most important result of this analysis is that the exponential evolution is a modification of the linear evolution of the solution. Therefore, the amount of deviation from the linear evolution can be controlled to preserve the positivity of the solution. Numerically, this procedure is equivalent to the correction of the residual of the numerical scheme corresponding to the linear evolution. Although being more complicated than the algorithm of [43], the present one can be applied to second-order implicit time-stepping schemes so that unsteady flows can be simulated more efficiently. One issue is that this correction procedure destroys the conservation of the numerical schemes. In the present paper, a second correction procedure is proposed to recover conservation. The proposed positivity-preserving algorithm is considerably less restrictive than existing ones. It does not rely on the existence of low-order positivity-preserving baseline schemes and convex decomposition of volume integrals. Furthermore, it can be implemented iteratively in the implicit dual time-stepping schemes to preserve positivity of the intermediate and converged states of the sub-iterations, without the need of the time-step size reduction. Therefore, the proposed algorithm is one of the most general positivity-preserving technique so far. It is proved that the present positivity-preserving algorithm is accuracy-preserving. Numerical results demonstrate that the proposed algorithm can preserve positivity of density and pressure when Euler and Navier-Stokes equations are solved using high-order FV schemes, and improves the robustness of the numerical scheme considerably.

The remainder of this paper is organized as follows. Numerical methods are introduced in Section 2. The asymptotic behaviors of the governing equations and the FV schemes near the points of local vacuum minimum will be discussed in Section 3. A general positivity-preserving algorithm for implicit high-order FV schemes solving Euler and Navier-Stokes equations will be proposed in Section 4. Numerical results are presented in Section 5 and concluding remarks are given in Section 6.

# 2. Numerical Methods

## 2.1 Governing equations

The three-dimensional non-dimensional compressible NS equations are in the form of

$$\frac{\partial \boldsymbol{U}}{\partial t} + \frac{\partial \boldsymbol{F}}{\partial x} + \frac{\partial \boldsymbol{G}}{\partial y} + \frac{\partial \boldsymbol{H}}{\partial z} = 0 \tag{1}$$

where $\boldsymbol{U}$ denote the conservative variables, $\boldsymbol{F}, \boldsymbol{G}$ and $\boldsymbol{H}$ represent the flux vectors. The specific forms of the conservative variables and flux terms are listed below.

$$U = (\rho, \rho u, \rho v, \rho w, \rho E)^T,$$
$$F = F_c - F_v, \quad G = G_c - G_v, \quad H = H_c - H_v$$
$$F_c = (\rho u, \rho u^2 + p, \rho uv, \rho uw, u(\rho E + p))^T,$$
$$G_c = (\rho v, \rho uv, \rho v^2 + p, \rho vw, v(\rho E + p))^T,$$
$$H_c = (\rho w, \rho uw, \rho vw, \rho w^2 + p, w(\rho E + p))^T,$$

$$F_v = \left(0, \tau_{xx}, \tau_{yx}, \tau_{zx}, u\tau_{xx} + v\tau_{yx} + w\tau_{zx} + k\frac{\partial T}{\partial x}\right)^T,$$
$$G_v = \left(0, \tau_{xy}, \tau_{yy}, \tau_{zy}, u\tau_{xy} + v\tau_{yy} + w\tau_{zy} + k\frac{\partial T}{\partial y}\right)^T,$$
$$H_v = \left(0, \tau_{xz}, \tau_{yz}, \tau_{zz}, u\tau_{xz} + v\tau_{yz} + w\tau_{zz} + k\frac{\partial T}{\partial z}\right)^T,$$

$$\tau_{xx} = 2\mu u_x + \lambda(u_x + v_y + w_z), \quad \tau_{xy} = \tau_{yx} = \mu(u_y + v_x),$$
$$\tau_{yy} = 2\mu v_y + \lambda(u_x + v_y + w_z), \quad \tau_{xz} = \tau_{zx} = \mu(u_z + w_x),$$
$$\tau_{zz} = 2\mu w_z + \lambda(u_x + v_y + w_z), \quad \tau_{yz} = \tau_{zy} = \mu(v_z + w_y).$$

In these formulations, the fluxes with the subscript "c" denote the inviscid fluxes and viscous fluxes are denoted by the subscript "v". $\rho$ denotes the density and $u$, $v$ and $w$ represent the components of the velocity vector. The total energy and pressure $p$ are related through

$$\rho E = \frac{p}{\gamma - 1} + \frac{1}{2}\rho(u^2 + v^2 + w^2), \tag{2}$$

where $\gamma$ denotes the specific heat ratio. In the formulations of the viscous stress terms such as $\tau_{xx}$ and $\tau_{xy}$, $\mu$ denotes the dynamic viscosity coefficient, and $\lambda$ is evaluated by $\lambda = -2/3\mu$ according to the Stokes hypothesis. In the energy equations, $T$ represents the temperature, and $k$ denotes the thermal conductivity which can be computed in terms of the viscosity $\mu$, the specific heat at constant pressure $c_p$, and the Prandtl number $\Pr$ through the relation

$$k = c_p \mu / \Pr.$$

It is worth noting that the density $\rho$ is a linear function of the conservative variables $U$ with

$$\rho(\theta_1 U_1 + \theta_2 U_2) = \theta_1 \rho(U_1) + \theta_2 \rho(U_2). \tag{3}$$

On the other hand, pressure $p$ is a concave function of the conservative variables $U$ satisfying

$$p(\theta_1 U_1 + \theta_2 U_2) \geq \theta_1 p(U_1) + \theta_2 p(U_2) \tag{4}$$

where the coefficients $\theta_1 + \theta_2 = 1$, $\theta_1 \geq 0$, $\theta_2 \geq 0$.

## 2.2 Spatial discretization

Dividing the computational domain $\Omega$ into $N_{cell}$ non-overlapping control volumes, i.e. $\Omega = \bigcup_{i=1}^{N_{cell}} \Omega_i$, and integrating Eq. (1) over a control volume $\Omega_i$, we have the integral forms of the governing equations

$$\frac{d\bar{U}_i}{dt} + \frac{1}{\bar{\Omega}_i} \oiint_{\partial \Omega_i} \vec{F} \cdot \vec{n} \, dS = \mathbf{0} \tag{5}$$

where

$$\bar{U}_i = \frac{1}{\bar{\Omega}_i} \iiint_{\Omega_i} U \, dV$$

is the cell average of the conservative variables,

$$\vec{F} = (F_c - F_v)\vec{i} + (G_c - G_v)\vec{j} + (H_c - H_v)\vec{k}$$

is the flux tensor, and $\vec{n}$ is the unit outward normal vector of $\partial \Omega_i$. For convenience, we denote Eq. (5) as

$$\frac{d\bar{U}_i}{dt} = -\mathbf{R}_i \tag{6}$$

with

$$\mathbf{R}_i = \frac{1}{\bar{\Omega}_i} \oiint_{\partial \Omega_i} \vec{F} \cdot \vec{n} \, dS . \tag{7}$$

In this paper, the cell-centered high-order FV methods will be used to discretize Eq. (6). Generally, a high-order FV scheme consists of three parts, i.e., reconstruction, flux evaluation, and time integration.

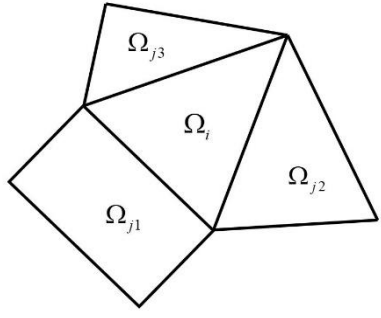

Fig. 1 The schematic diagram of the reconstruction stencil of cell $\Omega_i$: $S_i = \{\Omega_i, \Omega_{j_1}, \Omega_{j_2}, \Omega_{j_3}\}$

In the reconstruction step, on a control volume $\Omega_i$, a degree $k$ polynomial

$$u_i(\vec{x}) = \bar{u}_i + \boldsymbol{u}_i^T \boldsymbol{\psi}_i(\vec{x}) = \bar{u}_i + \sum_{l=1}^{N(k)} u_{i,l} \psi_{i,l}(\vec{x}) \tag{8}$$

for each component of the conservative variable vector denoted by $u$ is reconstructed from the cell averages $\bar{u}_j$ belonging to a reconstruction stencil, i.e., $\bar{u}_j \in S_i$. In Eq. (8), $\psi_{i,l}(\vec{x})$ denotes the Taylor basis function with zero mean, $u_{i,l}$ denotes the unknown coefficients of the reconstruction polynomial, and $N(k)$ denotes the number of unknown coefficients. For the two-dimensional situations, $N(k) = (k+1)(k+2)/2 - 1$, and for the three-dimensional situations, $N(k) = (k+1)(k+2)(k+2)/6 - 1$. In the present paper, the variational reconstruction (VR) proposed by Wang and Ren et al. [5] is adopted. The VR is an implicit reconstruction procedure with the reconstruction stencil being the entire domain. However, it can be designed so that all operations for the reconstruction of cell $\Omega_i$ are based on a compact stencil

$$S_i = \{\Omega_i, \Omega_{j_1}, \Omega_{j_2}, \Omega_{j_3}\} \tag{9}$$

shown in Fig.1, which includes only the cell of interest and it direct face-neighboring cells. For VR, the reconstruction procedure is derived by minimizing a cost function. The resulting linear system for solving the unknown coefficients is symmetric and positive definite so that a unique solution of the reconstruction problem can be guaranteed. This is the most significant advantage of VR over other reconstruction algorithms. Another advantage of this method is that it can achieve arbitrary order on accuracy on the operationally compact stencil. Generally speaking, the degree $k$ polynomial reconstruction is $k+1$-th order accurate if the solution is sufficiently smooth. The efficient implementation procedure of the reconstruction procedure can be also found in [5]. We should notice here that the positivity-preserving algorithm that will be proposed in the present paper is applicable to any reconstruction procedure. Therefore, we will not go into further details of this specific reconstruction algorithm.

In the flux evaluation step, the residual shown in Eq. (7) is computed by

$$R_i = \frac{1}{\bar{\Omega}_i} \sum_{f=1}^{N_f} \sum_{l=1}^{N_g} \omega_{f,l} J_{f,l} \left( \vec{F}_{f,l} \cdot \vec{n}_{f,l} \right) S_f \tag{10}$$

where $f$ denotes one of $N_f$ faces (or edges in 2D case) of the control volume $\Omega_i$. On each faces, the surface integration is carried out using the Gaussian quadrature rule with $\omega_{f,l}$ denoting the weight and $J_{f,l}$ denoting the Jacobian of the parametric space. On each Gaussian point, the numerical flux is denoted by

$$\vec{F}_{f,l} \cdot \vec{n}_{f,l} = F_{nc}\left( U_{f,l}^L, U_{f,l}^R, \vec{n}_{f,l} \right) - F_{nv}\left( U_{f,l}^L, U_{f,l}^R, \nabla U_{f,l}^L, \nabla U_{f,l}^R, \vec{n}_{f,l} \right)$$

where $F_{nc}$ is the inviscid flux, and $F_{nv}$ is the viscous flux. In this paper, the HLLE+ approximate Riemann solver [44] is used to evaluate of the inviscid fluxes, and the technique proposed by Wang et al. is used to compute viscous fluxes [5]. To facilitate the introduction of the positivity-preserving algorithm in the following section, Eq. (10) is written as

$$R_i = \frac{1}{\bar{\Omega}_i} \sum_{f=1}^{N_f} F_{n,f} \tag{11}$$

where

$$F_{n,f} = \sum_{l=1}^{N_g} \omega_{f,l} J_{f,l} \left( \vec{F}_{f,l} \cdot \vec{n}_{f,l} \right) S_f . \tag{12}$$

After the reconstruction and the flux evaluation procedures, the temporal discretization is carried out to advance the numerical solution in time. Since the present positivity-preserving algorithm is applied with an implicit temporal scheme, the temporal scheme will be introduced in detail in the next sub-section.

## 2.3 Time integration

In the present paper, the second-order backward difference (BDF2) implicit time integration scheme is adopted for the temporal discretization. When solving the unsteady problem, the dual time-stepping method is applied. The sub-iteration in terms of the pseudo time is carried out using the LU-SGS method [45]. The BDF2 implicit time integration scheme with the pseudo-temporal term can be expressed as

$$\frac{\partial \bar{U}_i^{n+1}}{\partial \tau} + \frac{(1+\phi)(\bar{U}_i^{n+1}-\bar{U}_i^n)-\phi(\bar{U}_i^n-\bar{U}_i^{n-1})}{\Delta t} = -R_i^{n+1} \tag{13}$$

where $\tau$ is the pseudo time, $\Delta t$ denotes the physical time step, and the parameter $\phi$ is chosen as $\phi = 1/2$ for the temporally second-order scheme. A first-order backward difference scheme is used to discretize the pseudo-temporal term, i.e.,

$$\frac{\bar{U}_i^{n+1,m+1}-\bar{U}_i^{n+1,m}}{\Delta \tau} + \frac{(1+\phi)(\bar{U}_i^{n+1,m+1}-\bar{U}_i^n)-\phi(\bar{U}_i^n-\bar{U}_i^{n-1})}{\Delta t} = -R_i^{n+1,m+1} \tag{14}$$

where the superscript $m$ is the index of pseudo time step of the sub-iteration. The spatial term $R_i^{n+1,m+1}$ in Eq. (14) is linearized at $\tau_m$ so that a linear system is resulted for the sub-iteration. Then, the LU-SGS method is applied to solve this system of linear equations. The forward and backward sweepings of the LU-SGS method are respectively,

$$\Delta \bar{U}_i^{n+1,m,*} = \left[ -\widehat{R}_i^{n+1,m} - \frac{1}{\bar{\Omega}_i} \sum_{\Omega_j \in S_i, j<i} \frac{S_{i,j}}{2}\left(\frac{\partial F_n}{\partial U_j}-\lambda_{ij}\right)\Delta\bar{U}_j^{n+1,m,*} \right] \bigg/ \left( \frac{1}{\Delta t} + \frac{1}{\Delta \tau_i} + \frac{1}{\bar{\Omega}_i}\sum_{\Omega_j \in S_i}\frac{S_{i,j}}{2}\lambda_{ij} \right)$$

$$\Delta \bar{U}_i^{n+1,m} = \Delta \bar{U}_i^{n+1,m,*} - \frac{1}{\bar{\Omega}_i}\sum_{\Omega_j \in S_i, j>i}\frac{S_{i,j}}{2}\left(\frac{\partial F_n}{\partial U_j}-\lambda_{ij}\right)\Delta\bar{U}_j^{n+1,m} \bigg/ \left( \frac{1}{\Delta t} + \frac{1}{\Delta \tau_i} + \frac{1}{\bar{\Omega}_i}\sum_{\Omega_j \in S_i}\frac{S_{i,j}}{2}\lambda_{ij} \right)$$

$$\tag{15}$$

where

$$\widehat{R}_i^{n+1,m} = R_i^{n+1,m} + \frac{(1+\phi)(\bar{U}_i^{n+1,m}-\bar{U}_i^n)-\phi(\bar{U}_i^n-\bar{U}_i^{n-1})}{\Delta t} \tag{16}$$

is the residual of the dual time-stepping scheme. The local pseudo time-stepping technique is adopted to enhance the convergence speed of pseudo time sub-iteration. After solving Eq. (15), the conservative variables are updated through,

$$\bar{U}_i^{n+1,m+1} = \bar{U}_i^{n+1,m} + \Delta \bar{U}_i^{n+1,m}. \tag{17}$$

Supposing sub-iteration is converged after $M$ pseudo time steps, we obtained the numerical solution at $t_{n+1}$, i.e.

$$\bar{U}_i^{n+1} = \bar{U}_i^{n+1,M+1}. \tag{18}$$

# 3. The asymptotic behaviors of the governing equations and the finite volume schemes

## 3.1 Asymptotic behaviors of the Euler and NS equations near the points of local vacuum minimum

Since the design of the positivity-preserving algorithms are mathematically difficult for implicit temporal scheme solving Euler and NS equations, it is helpful to study the asymptotic behaviors of the governing equations at locations where the non-positive numerical solutions are likely to occur. This physical-based analysis will be used to guide the design of the positivity-preserving algorithm of the present paper.

In the numerical computation, there are two situations in which the negative density or pressure may occur. The first one is near the shock or contact waves where the non-physical oscillations in the numerical solution will lead to local negative density or pressure at locations already with small density or pressure. The second one is in the strong expansion waves with solutions close to the vacuum states, where numerical errors will produce negative density or pressure locally. In both cases, the negative density or pressure will occur firstly at local minimum points of density or pressure.

An imaginary state called the point of local vacuum minimum (PLVM) can thus be defined.

**Definition**. If there exists a point $\vec{x}_0$, such that

$$\begin{aligned} \rho(\vec{x}_0,t) &= 0, \quad \nabla \rho(\vec{x}_0,t) = \vec{0}, \\ p(\vec{x}_0,t) &= 0, \quad \nabla p(\vec{x}_0,t) = \vec{0}, \end{aligned} \tag{19}$$

then $\vec{x}_0$ is called a PLVM.

In what follows, we will show some properties that can be inferred from the definition of the PLVM.

**Property 1**. The existence of the PLVM. In the numerical or analytical solutions of the compressible flows, if there is one isolate point $\vec{x}_0$ where $\rho(\vec{x}_0,t)=0$ or $p(\vec{x}_0,t)=0$, this point is a PLVM.

For the point where $\rho(\vec{x}_0,t)=0$, there must be $\rho(\vec{x},t)>0$ in its neighborhood. If the solution is smooth, the point is a local minimum and thus $\nabla\rho(\vec{x}_0,t)=\vec{0}$. For perfect gases, the equation of state is

$$p = \rho RT.$$

This indicates $p(\vec{x}_0,t)=0$ and the neighborhood pressure satisfies $p(\vec{x},t)>0$ if the absolute temperature is not zero. This again leads to $\nabla p(\vec{x}_0,t)=\vec{0}$. Therefore, this point is a PLVM. For the case of $p(\vec{x}_0,t)=0$, the same conclusion can be drawn.

This property shows that the PLVM is a critical point before the negative density or pressure occurs in the numerical simulation. It is thus important to study what will happen in the neighborhood of this point.

**Property 2.** In the neighborhood of a PLVM $\vec{x}_0$, the density, pressure and their gradients are small quantities.

Using the Taylor series expansion, we notice that the density and its gradients can be expanded as

$$\rho(\vec{x},t) = \rho(\vec{x}_0,t) + \nabla\rho(\vec{x},t)\big|_{\vec{x}=\vec{x}_0} \cdot (\vec{x}-\vec{x}_0) + (\vec{x}-\vec{x}_0)\cdot\nabla\nabla\rho(\vec{x},t)\big|_{\vec{x}=\vec{x}_0} \cdot (\vec{x}-\vec{x}_0) + \cdots$$
$$= (\vec{x}-\vec{x}_0)\cdot\nabla\nabla\rho(\vec{x},t)\big|_{\vec{x}=\vec{x}_0} \cdot (\vec{x}-\vec{x}_0) \sim h^2$$
$$\nabla\rho(\vec{x},t) = \nabla\rho(\vec{x},t)\big|_{\vec{x}=\vec{x}_0} + \nabla\nabla\rho(\vec{x},t)\big|_{\vec{x}=\vec{x}_0} \cdot (\vec{x}-\vec{x}_0) + \cdots$$
$$= \nabla\nabla\rho(\vec{x},t)\big|_{\vec{x}=\vec{x}_0} \cdot (\vec{x}-\vec{x}_0) \sim h$$

where $h$ is the size of the neighborhood of $\vec{x}_0$. The pressure has similar property. Therefore, in the neighborhood of a PLVM, both density and pressure are small and slow varying.

**Property 3**. The flow is in expansion in the neighborhood of a PLVM.

Considering the continuity equation

$$\frac{D\rho}{Dt} + \rho \nabla \cdot \vec{u} = 0$$

If the numerical result of density changes from positive to negative during time interval $t_n$ and $t_{n+1}$, then it obvious that

$$\frac{D\rho}{Dt} < 0$$

Thus, we have

$$\nabla \cdot \vec{u} > 0 \tag{20}$$

Therefore, Eq.(19) corresponds to an expansion process.

In what follows, we will analyze the temporally asymptotic behavior of the governing equation near the PLVM. To simplify the discussion, we analyze the Euler equations, and the extension to the NS equations will be discussed after this analysis.

The vector form of the Euler equations can be written as

$$\begin{aligned}
&\frac{\partial \rho}{\partial t} + \rho \nabla \cdot \vec{u} = -\vec{u} \cdot \nabla \rho, \\
&\frac{\partial \rho \vec{u}}{\partial t} + \rho \vec{u} \nabla \cdot \vec{u} = -\left(\vec{u}(\vec{u} \cdot \nabla \rho) + \rho \vec{u} \cdot \nabla \vec{u} + \nabla p\right), \\
&\frac{\partial \rho E}{\partial t} + \rho E \nabla \cdot \vec{u} = -\left(p \nabla \cdot \vec{u} + \frac{\gamma}{\gamma - 1} \vec{u} \cdot \nabla p + \frac{1}{2}(\vec{u} \cdot \vec{u}) \vec{u} \cdot \nabla \rho + \frac{\rho \vec{u}}{2} \cdot \nabla (\vec{u} \cdot \vec{u})\right).
\end{aligned} \tag{21}$$

Eq. (21) shows that in the neighborhood of a PLVM, the evolution of the dependent variables is mainly driven by expansion in terms of $\nabla \cdot \vec{u}$ according to property 3. Furthermore, the right-hand sides of Eq. (21) are all small quantities according to properties 1 and 2. Therefore, it is reasonable to assume that the right-hand sides of Eq. (21) can be frozen to constant states both temporally and spatially. If we further assume that the expansion is uniform, Eq. (21) can be simplified to

$$\frac{\partial \rho}{\partial t} + \rho \tilde{D} = -\tilde{S}_\rho,$$

$$\frac{\partial \rho u}{\partial t} + \rho u \tilde{D} = -\tilde{S}_{\rho u},$$

$$\frac{\partial \rho v}{\partial t} + \rho v \tilde{D} = -\tilde{S}_{\rho v}, \qquad (22)$$

$$\frac{\partial \rho w}{\partial t} + \rho w \tilde{D} = -\tilde{S}_{\rho w},$$

$$\frac{\partial \rho E}{\partial t} + \rho E \tilde{D} = -\tilde{S}_{\rho E},$$

where

$$\tilde{D} = \nabla \cdot \vec{u} > 0$$

can be considered as the rate of expansion,

$$\tilde{S}_\rho = \vec{u} \cdot \nabla \rho,$$

$$\tilde{S}_{\rho u} = \rho \vec{u} \cdot \nabla u + u(\vec{u} \cdot \nabla \rho) + \frac{\partial p}{\partial x},$$

$$\tilde{S}_{\rho v} = \rho \vec{u} \cdot \nabla v + v(\vec{u} \cdot \nabla \rho) + \frac{\partial p}{\partial y},$$

$$\tilde{S}_{\rho w} = \rho \vec{u} \cdot \nabla w + w(\vec{u} \cdot \nabla \rho) + \frac{\partial p}{\partial z},$$

$$\tilde{S}_{\rho E} = p \nabla \cdot \vec{u} + \frac{\gamma}{\gamma - 1} \vec{u} \cdot \nabla p + \frac{1}{2}(\vec{u} \cdot \vec{u}) \vec{u} \cdot \nabla \rho + \frac{\rho \vec{u}}{2} \cdot \nabla(\vec{u} \cdot \vec{u}),$$

are the forcing terms, and the tilde symbols denote the constant frozen states. Eq. (22) is a system of ordinary equations. Supposing $\psi$ is one of the dependent variables, we can re-write Eq. (22) as

$$\frac{d\psi}{dt} + \psi \tilde{D} = -\tilde{S}_\psi. \qquad (23)$$

Given the initial condition at $t_n$, the solution of Eq. (23) at $t \geq t_n$ is

$$\psi(t) + \frac{\tilde{S}_\psi}{\tilde{D}} = \left(\psi^n + \frac{\tilde{S}_\psi}{\tilde{D}}\right) e^{-\tilde{D}(t - t_n)}. \qquad (24)$$

At a PLVM, it is straightforward to show that $\tilde{S}_\psi = 0$. Eq. (24) reduces to

$$\psi(t) = \psi^n e^{-\tilde{D}(t - t_n)}. \qquad (25)$$

Therefore, the solution is in exponentially decay at this PLVM. Furthermore, at this point, since

$\rho^n = 0$ and $p^n = 0$, we have in fact $\psi^n = 0$. The actual solution is thus

$$\psi(t) = 0 \tag{26}$$

This guarantees that the density and pressure are non-negative. This result indicates that the analytical solutions can never predict negative pressure or density if the solutions are smooth. This is because before negative pressure or density appears, the solutions have to pass through the PLVM states. Then the solutions are trapped at the zero solutions of Eq. (26).

However, in the numerical simulations, the solutions are discrete in time. The state of Eq. (26) cannot be reached numerically in most situations. Therefore, it is important to discuss the positivity-preserving mechanism of the asymptotic solutions in the vicinity of a PLVM to provide a guidance for the design of the positivity-preserving algorithms.

Practically, it is sufficient to ask the density and pressure is non-negative for $t_n \leq t \leq t_n + \Delta t$ where $\Delta t$ is the physical time step of the numerical simulation. We assume that $\psi^n \neq 0$ near the PLVM. For density, we have

$$\rho(t) + \frac{\tilde{S}_\rho}{\tilde{D}} = \left(\rho^n + \frac{\tilde{S}_\rho}{\tilde{D}}\right) e^{-\tilde{D}(t-t_n)}. \tag{27}$$

It is straightforward to show that

$$\left(\rho^n + \frac{\tilde{S}_\rho}{\tilde{D}}\right) e^{-\tilde{D}(t-t_n)} - \frac{\tilde{S}_\rho}{\tilde{D}} \geq 0,$$

or

$$\frac{\tilde{S}_\rho}{\tilde{D}} < \frac{\rho^n e^{-\tilde{D}\Delta t}}{1 - e^{-\tilde{D}\Delta t}} \tag{28}$$

is sufficient for the density to be non-negative at $t = t_n + \Delta t$. When $\tilde{S}_\rho \leq 0$, Eq.(28) will always be satisfied. Therefore, a stricter condition is

$$\left|\frac{\tilde{S}_\rho}{\tilde{D}}\right| < \frac{\rho^n e^{-\tilde{D}\Delta t}}{1 - e^{-\tilde{D}\Delta t}}. \tag{29}$$

Now we consider the pressure. Substituting Eq.(24) into Eq.(2), we have

$$p(t) = (\gamma-1) \left\{ \begin{array}{l} (\rho E)^n e^{-\tilde{D}(t-t_n)} - \dfrac{\tilde{S}_{\rho E}}{\tilde{D}}\left(1-e^{-\tilde{D}(t-t_n)}\right) \\[2ex] - \dfrac{\left[(\rho u)^n e^{-\tilde{D}(t-t_n)} - \dfrac{\tilde{S}_{\rho u}}{\tilde{D}}\left(1-e^{-\tilde{D}(t-t_n)}\right)\right]^2}{2\left[\rho^n e^{-\tilde{D}(t-t_n)} - \dfrac{\tilde{S}_{\rho}}{\tilde{D}}\left(1-e^{-\tilde{D}(t-t_n)}\right)\right]} \\[3ex] - \dfrac{\left[(\rho v)^n e^{-\tilde{D}(t-t_n)} - \dfrac{\tilde{S}_{\rho v}}{\tilde{D}}\left(1-e^{-\tilde{D}(t-t_n)}\right)\right]^2}{2\left[\rho^n e^{-\tilde{D}(t-t_n)} - \dfrac{\tilde{S}_{\rho}}{\tilde{D}}\left(1-e^{-\tilde{D}(t-t_n)}\right)\right]} \\[3ex] - \dfrac{\left[(\rho w)^n e^{-\tilde{D}(t-t_n)} - \dfrac{\tilde{S}_{\rho w}}{\tilde{D}}\left(1-e^{-\tilde{D}(t-t_n)}\right)\right]^2}{2\left[\rho^n e^{-\tilde{D}(t-t_n)} - \dfrac{\tilde{S}_{\rho}}{\tilde{D}}\left(1-e^{-\tilde{D}(t-t_n)}\right)\right]} \end{array} \right\}. \tag{30}$$

It can be shown that there exists a coefficient $0 < \xi \leq 1$, if the following condition is satisfied for all $\psi$ near a PLVM

$$\left|\dfrac{\tilde{S}_\psi}{\tilde{D}}\right| \leq \xi |\psi^n| \dfrac{e^{-\tilde{D}\Delta t}}{1-e^{-\tilde{D}\Delta t}} \tag{31}$$

then the pressure is not negative in a neighborhood of this PLVM at $t=t_n+\Delta t$. We omit the proof for brevity. Since Eq.(31) is stricter than Eq.(29), it also guarantees the positivity of density $\rho^{n+1}$.

This condition indicates that the main mechanism for the positivity-preserving property of the Euler equations is that the forcing term $\left|\tilde{S}_\psi\right|$ should be sufficiently small. This is partially supported by the property 2 of the PLVM. Therefore, in designing the numerical positivity-preserving algorithms, a proper control of $\left|\tilde{S}_\psi\right|$ is a plausible measurement to preserve the positivity of the density and pressure.

The above discussion is also applicable to the NS equations. Because the density is small near the PLVM, the dynamic viscosity coefficient is also small. Therefore, the viscous terms can be treated as small and slow varying quantities. They can therefore be frozen to constant states so that the discussion for the Euler equations is still valid for the NS equations.

## 3.2 Asymptotic behaviors of the discretized equations near the PLVM

According to the above discussion, Euler and NS equations exhibit an approximate exponential time evolution property near the PLVM, which is the basis for preserving the positivity of the density and pressure during the physical process. The following discussion will demonstrate that this property can be also used to design a numerically positivity-preserving scheme. For the numerical schemes, only the FV method is considered in this paper. However, the basic ideas can be also applied to other methods.

Eq. (23) can be written as

$$\frac{d}{dt}\boldsymbol{U} + \boldsymbol{U}\tilde{D} + \tilde{\boldsymbol{S}} = \boldsymbol{0}, \tag{32}$$

where

$$\tilde{D} = \nabla \cdot \bar{u}, \\ \tilde{\boldsymbol{S}} = \left(\tilde{S}_\rho, \tilde{S}_{\rho u}, \tilde{S}_{\rho v}, \tilde{S}_{\rho w}, \tilde{S}_{\rho E}\right)^T. \tag{33}$$

Integrating Eq. (32) over a control volume $\Omega_i$, the following FV discretization

$$\frac{d}{dt}\bar{\boldsymbol{U}}_i + \bar{\boldsymbol{U}}_i \tilde{D}_i + \tilde{\boldsymbol{S}}_i = \boldsymbol{0} \tag{34}$$

is resulted near the PLVM. Eq. (34) is a system of ordinary differential equations. By integrating Eq. (34) from $t_n$ to $t_{n+1}$, we have

$$\bar{\boldsymbol{U}}_i^{n+1} = \bar{\boldsymbol{U}}_i^n + \left(\bar{\boldsymbol{U}}_i^n + \frac{\tilde{\boldsymbol{S}}_i}{\tilde{D}_i}\right)\left(e^{-\tilde{D}_i \Delta t} - 1\right). \tag{35}$$

Eq. (35) usually cannot guarantees the positivity of the density and pressure. According to the analysis of Section 3.1, to preserve the positivity, the absolute values of the components of $\tilde{\boldsymbol{S}}_i$ should be sufficiently small. According to Eq. (34), at $t_n$,

$$\tilde{\boldsymbol{S}}_i = -\left(\frac{d\bar{\boldsymbol{U}}}{dt}\right)_i^n - \tilde{D}_i \bar{\boldsymbol{U}}_i^n. \tag{36}$$

The control of $\tilde{\boldsymbol{S}}_i$ is equivalent to the control of $\left(d\bar{\boldsymbol{U}}/dt\right)_i^n$ or $\tilde{D}_i \bar{\boldsymbol{U}}_i^n$. Using the continuity equation as an example, for which

$$\tilde{S}_\rho = -\left(\frac{d\rho}{dt}\right)^n - \rho^n \tilde{D}.$$

In the neighborhood of a PLVM, there will be $(d\rho/dt)^n < 0$, $\rho^n \tilde{D} > 0$, and the negative density can occur only when $\tilde{S}_\rho > 0$. In this case, it is necessary to reduce the magnitude of $(d\rho/dt)^n$ instead of $\rho^n \tilde{D}$ since it is $(d\rho/dt)^n$ that makes the density being negative. If this idea is also applied to other variables, Eq.(36) becomes

$$\tilde{S}_i = -\xi_i \left(\frac{d\bar{U}}{dt}\right)^n_i - \tilde{D}_i \bar{U}^n_i \tag{37}$$

where $0 < \xi_i \leq 1$. Substituting Eq.(37) into Eq.(35), we have

$$\bar{U}_i^{n+1} - \bar{U}_i^n = \alpha_i \left(\frac{d\bar{U}}{dt}\right)^n_i \Delta t \tag{38}$$

where

$$\alpha_i = \xi_i \theta_i, \quad \theta_i = \frac{\left(1 - e^{-\tilde{D}_i \Delta t}\right)}{\tilde{D}_i \Delta t}.$$

Since $\tilde{D}_i$ and $\tilde{S}_i$ are frozen states in Eq. (36),

$$\left(\frac{d\bar{U}}{dt}\right)^n_i$$

is in fact a constant corresponding to a linear evolution, i.e.

$$\left(\frac{d\bar{U}}{dt}\right)^n_i = \frac{\left(\bar{U}_i^{n+1} - \bar{U}_i^n\right)_{\text{linear}}}{\Delta t}.$$

In order to distinguish from the linear evolution, the left-hand side of Eq.(38) is termed as

$$\left(\bar{U}_i^{n+1} - \bar{U}_i^n\right)_{\text{exponential}}$$

to stand for the actual exponential evolution. Eq.(38) can thus be written as

$$\left(\bar{U}_i^{n+1} - \bar{U}_i^n\right)_{\text{exponential}} = \alpha_i \left(\bar{U}_i^{n+1} - \bar{U}_i^n\right)_{\text{linear}}. \tag{39}$$

Eq. (39) shows that the asymptotic solution with exponential evolution can be treated as the modification of the linear evolution process. The correction factor $\alpha_i = \xi_i \theta_i$ is determined by two processes: the first one is the exponential evolution itself, the second one is due to the modification of the forcing term $\tilde{S}_i$. In designing the positivity-preserving schemes, we do not need to distinguish these two processes and only need to determine $\alpha_i$ directly.

Next, we will discuss how should Eq. (39) be used to develop the present positivity-preserving algorithm. As having been introduced in Section 2.3, we use BDF2 as the time discretization scheme, and the fully discretized FV scheme is

$$\frac{(1+\phi)\left(\bar{U}_i^{n+1}-\bar{U}_i^{n}\right)-\phi\left(\bar{U}_i^{n}-\bar{U}_i^{n-1}\right)}{\Delta t}=-R_i^{n+1}. \tag{40}$$

To achieve second-order accuracy, Eq. (40) assumes that a temporally quadratic evolution in the numerical solution. Since exponential evolution is a modification of linear evolution, it is necessary to separate the linear evolution part of Eq. (40) with other terms in order to apply Eq.(39). For this purpose, we re-write Eq. (40) as

$$\bar{U}_i^{n+1}-\bar{U}_i^{n}=\frac{\Delta t}{1+\phi}\left(-R_i^{n+1}+\phi\frac{\bar{U}_i^{n}-\bar{U}_i^{n-1}}{\Delta t}\right). \tag{41}$$

We denote

$$\tilde{R}_i^{n+1}=\frac{1}{1+\phi}\left(R_i^{n+1}-\phi\frac{\bar{U}_i^{n}-\bar{U}_i^{n-1}}{\Delta t}\right), \tag{42}$$

which is called the linear residual corresponding to the linear evolution process. Eq. (41) can thus be denoted as

$$\bar{U}_i^{n+1}=\bar{U}_i^{n}-\tilde{R}_i^{n+1}\Delta t \tag{43}$$

or

$$\left(\bar{U}_i^{n+1}-\bar{U}_i^{n}\right)_{\text{linear}}=-\tilde{R}_i^{n+1}\Delta t.$$

The application of Eq. (39) leads to

$$\left(\bar{U}_i^{n+1}-\bar{U}_i^{n}\right)_{\text{exponential}}=\alpha_i\left(\bar{U}_i^{n+1}-\bar{U}_i^{n}\right)_{\text{linear}}=-\alpha_i\tilde{R}_i^{n+1}\Delta t. \tag{44}$$

Eq. (44) shows that, by incorporating the exponential evolution modification, the FV scheme is changed to

$$\bar{U}_i^{n+1}=\bar{U}_i^{n}-\alpha_i\tilde{R}_i^{n+1}\Delta t \tag{45}$$

near the PLVM.

It is clear that for sufficiently small $\alpha_i$, the solution is positivity-preserving. In Section 4, the procedures to determine $\alpha_i$ will be discussed in detail. It can be also proved that the determined $\alpha_i$ is very close to 1, and the order of accuracy of the numerical scheme is not affected. One issue should be noticed: the direct application of Eq. (45) will destroy the conservation of the numerical scheme. In Section 4, the procedure to recover the conservation will also be introduced.

Finding $\alpha_i$ to preserve the positivity is easier than other approaches in the literature mathematically. Furthermore, as long as the linear evolution residual can be identified, the current method is independent of the spatial schemes, and does not rely on the positivity-preserving low-order baseline schemes. Therefore, it is possible to design a general positivity-preserving algorithm applicable to high-order FV schemes solving Euler and NS equations with implicit temporal discretizations.

# 4 Positivity-preserving algorithms for the high-order implicit FV schemes

## 4.1 General considerations

The positivity-preserving problem is stated as follows. Defining the set of positivity-preserving solutions as [28]

$$U \in G = \left\{ (\rho, \rho \vec{u}, \rho E) \big| \rho > 0, p > 0 \right\}, \tag{46}$$

we design a numerical scheme whose solution at $t_{n+1}$ satisfies $U^{n+1} \in G$ if $U^n \in G$. The solution is possible with zero pressure or density, however, to avoid possible numerical difficulties, we ask $\rho > 0$ and $p > 0$ in the present paper.

When the high-order FV schemes are applied in the numerical simulation, the positivity-preserving scheme should be subject to the following requirement. On any given control volume $\Omega_i$, $U_i^{n+1}(\vec{x}) \in G$, for $\forall \vec{x} \in \Omega_i$, where

$$U_i^{n+1}(\bar{x}) = \bar{U}_i^{n+1} + \left(u_i^{n+1}\right)^T \psi_i(\bar{x}) \tag{47}$$

is the polynomial reconstruction of the numerical solution. To fully comply with this requirement, one should find the minimum pressure or density determined by Eq. (47), which demands considerable efforts in the numerical computation. Practically, it is sufficient to ask

$$U_i^{n+1}(\bar{x}_j) \in G \tag{48}$$

at some characteristic points $x_j \in P_i \subset \Omega_i$ [28]. In the present paper, $P_i$ consists of all Gaussian quadrature points of the volume integration which is sufficient for computing physically consistent cell-averaged quantities, and consists of all Gaussian quadrature points of surface integration on the cell interfaces which is sufficient for computing physically correct numerical fluxes.

To design such a positivity-preserving algorithms, the following two-step approach is adopted in this paper. In the first step, $\alpha_i$ in Eq. (45) is determined to ensure $\bar{U}_i^{n+1}$ is positivity-preserving, i.e.

$$\bar{U}_i^{n+1} \in \bar{G} = \left\{ \left(\bar{\rho}, \overline{\rho u}, \overline{\rho E}\right) \middle| \bar{\rho} > 0, \bar{p} > 0 \right\}. \tag{49}$$

$\bar{U}_i^{n+1}$ is known only after the sub-iteration of the dual time-stepping approach mentioned in Section 2.3 is convergent. To solve this problem, the positivity-preserving algorithm is applied in every sub-iteration. The idea is that if $\bar{U}_i^{n+1,m} \in \bar{G}$ in every sub-iteration $m$, we have $\bar{U}_i^{n+1} \in \bar{G}$ after the iteration converges.

In the second step, the reconstruction is carried out to compute the reconstruction polynomial

$$U_i^{n+1}(\bar{x}) = \bar{U}_i^{n+1} + \left(u_i^{n+1}\right)^T \psi_i(\bar{x}). \tag{50}$$

Then a positivity-preserving limiter $\beta_i$ is applied to Eq. (50) resulting

$$U_i^{n+1}(\bar{x}) = \bar{U}_i^{n+1} + \beta_i \left(u_i^{n+1}\right)^T \psi_i(\bar{x}). \tag{51}$$

The procedure of Zhang and Shu [28] is used to determine $\beta_i$ so that Eq. (48) is satisfied for $x_j \in P_i \subset \Omega_i$. The detailed algorithm to determine $\beta_i$ can be referred to [28]. It will not be repeated here.

It is worth noting that in the algorithm of [28], the procedure to determine $\beta_i$ is the only measurement to enforce the positivity of the numerical solution. It is coupled with the flux functions and is relied the strong stability preserving (SSP) property of the temporal discretization scheme. Therefore, the proposed procedure is very complicated, and cannot be used for implicit temporal schemes. Even for the explicit SSP schemes, it is necessary to reduce the time step to maintain the stability of the numerical scheme. On the other hand, in the present paper, the procedure of Eq. (51) is only to guarantee the volume integrations and numerical fluxes can be computed correctly, and the positivity of the cell average solutions has already been achieved in the first step. Therefore, the present approach is much simpler than that of [28]. Furthermore, the second step is independent of the first step, which will not lead to the reduction of the time step and can be used for implicit time marching schemes.

In what follows, only the first step of the current positivity-preserving procedure to achieve Eq. (49) will be discussed.

## 4.2 Determination of cells with possible negative density or pressure

The present positivity-preserving algorithm is only applied to cells with possible negative density or pressure. The numerical solution is labeled as

$$\bar{U}_i^{n+1} \notin \bar{G}$$

if the numerical solution $\bar{U}_i^{n+1}$ on cell $\Omega_i$ has possible negative density or pressure, which is denoted as

$$\bar{\rho}_i^{n+1} \notin \bar{G}$$

or

$$\bar{p}_i^{n+1} \notin \bar{G}$$

respectively. Ideally, the positivity of $\bar{U}_i^{n+1}$ is known only after the sub-iteration of the dual time-stepping approach mentioned in Section 2.3 is convergent. However, since the present positivity-preserving algorithm is implemented in every sub-iteration step, the positivity of the solution can be estimated using

$$\bar{U}_i^{n+1,m+1,*} = \bar{U}_i^n - \tilde{R}_i^{n+1,m}\Delta t . \qquad (52)$$

where the superscript * stands for the estimated solution when the correction factor is not applied,

i.e $\alpha_i = 1$ in Eq. (45).

Firstly, we estimate the positivity of density. Considering the first component of Eq. (52), i.e.,

$$\bar{\rho}_i^{n+1,m+1,*} = \bar{\rho}_i^n - \tilde{R}_{i,\rho}^{n+1,m} \Delta t \quad (53)$$

we could check the positivity of density by the condition

$$\bar{\rho}_i^{n+1,m+1,*} = \bar{\rho}_i^n - \tilde{R}_{i,\rho}^{n+1,m} \Delta t < 0.$$

However, since $\bar{\rho}_i^{n+1,m+1,*}$ is not the final solution, a more conservative estimation is to check if

$$\bar{\rho}_i^{n+1,m+1,*} = \bar{\rho}_i^n - \tilde{R}_{i,\rho}^{n+1,m} \Delta t < \delta_\rho \quad (54)$$

where the parameter $\delta_\rho$ is chosen as $\delta_\rho = 10^{-15}$. The numerical tests show that, with the tolerance $\delta_\rho$, cells with possible negative density can be detected entirely.

Secondly, we estimate the positivity of pressure. According to Eq. (52), the estimated cell average of the pressure is

$$\bar{p}_i^{n+1,m+1,*} = (\gamma - 1) \left\{ \left( \overline{\rho E}_i^n - \tilde{R}_{i,\rho E}^{n+1,m} \Delta t \right) - \frac{1}{2\Omega} \iiint_\Omega \left[ \frac{\left( \overline{\rho u}_i^n - \tilde{R}_{i,\rho u}^{n+1,m} \Delta t + \beta_i^m \delta(\rho u)_i^{n+1,m} \right)^2}{\bar{\rho}_i^n - \tilde{R}_{i,\rho}^{n+1,m} \Delta t + \beta_i^m \delta \rho_i^{n+1,m}} + \frac{\left( \overline{\rho v}_i^n - \tilde{R}_{i,\rho v}^{n+1,m} \Delta t + \beta_i^m \delta(\rho v)_i^{n+1,m} \right)^2}{\bar{\rho}_i^n - \tilde{R}_{i,\rho}^{n+1,m} \Delta t + \beta_i^m \delta \rho_i^{n+1,m}} + \frac{\left( \overline{\rho w}_i^n - \tilde{R}_{i,\rho w}^{n+1,m} \Delta t + \beta_i^m \delta(\rho w)_i^{n+1,m} \right)^2}{\bar{\rho}_i^n - \tilde{R}_{i,\rho}^{n+1,m} \Delta t + \beta_i^m \delta \rho_i^{n+1,m}} \right] dxdydz \right\} \quad (55)$$

where $\delta(\bullet)$ denotes the difference between the reconstructed polynomials and the corresponding cell averaged values by writing Eq. (47) as

$$\boldsymbol{U}_i^{n+1}(\vec{x}) = \bar{\boldsymbol{U}}_i^{n+1} + \delta\left(\boldsymbol{U}_i^{n+1}\right).$$

Specifically, we have

$$\delta \rho_i = \boldsymbol{u}_{i,1}^T \boldsymbol{\psi}_i(\vec{x}), \quad \delta(\rho u)_i = \boldsymbol{u}_{i,2}^T \boldsymbol{\psi}_i(\vec{x}),$$
$$\delta(\rho v)_i = \boldsymbol{u}_{i,3}^T \boldsymbol{\psi}_i(\vec{x}), \quad \delta(\rho w)_i = \boldsymbol{u}_{i,4}^T \boldsymbol{\psi}_i(\vec{x}).$$

The $\beta_i^m$ in Eq. (55) denotes the positivity-preserving limiter of Eq. (51) at iteration step $m$. If

$$\overline{p}_i^{n+1,m+1,*} < \delta_p, \tag{56}$$

then the pressure is possibly negative. Similar to density, we choose $\delta_p = 10^{-15}$. If Eq. (54) or Eq. (56) is valid, then the solution at the corresponding cell is labeled as

$$\overline{\rho}_i^{n+1,m+1} \notin \overline{G}$$

or

$$\overline{p}_i^{n+1,m+1} \notin \overline{G},$$

either of which will lead to

$$\overline{U}_i^{n+1,m+1} \notin \overline{G}.$$

We notice the evaluation of Eq. (55) is time-consuming. To reduce the cost of computation, we can firstly compute the second-order estimation of the cell average of the pressure using

$$\overline{p}^{n+1,m+1,**} \approx (\gamma-1)\left[\overline{\rho E}^{n+1.m+1,*} - \frac{\left(\overline{\rho u}^{n+1.m+1,*}\right)^2 + \left(\overline{\rho v}^{n+1.m+1,*}\right)^2 + \left(\overline{\rho w}^{n+1.m+1,*}\right)^2}{2\overline{\rho}^{n+1}}\right]. \tag{57}$$

If

$$\overline{p}^{n+1,m+1,**} < 2\delta_p$$

we will use Eq. (55) to computed a refined estimation of the pressure. Otherwise, the pressure is treated to be positive.

## 4.3 The linear residual correction

If

$$\overline{U}_i^{n+1,m+1} \notin \overline{G}$$

we will determine the correction factor $\alpha_i$ in Eq. (45) to enforce the corrected $\overline{U}_i^{n+1,m+1}$ satisfies

$$\overline{U}_i^{n+1,m+1} \in \overline{G}.$$

We notice that this is always possible since if $\alpha_i \to 0$, we have

$$\overline{U}^{n+1,m+1} \to \overline{U}^n,$$

where $\overline{U}^n$ is always assumed to be positivity-preserving. The application of $\alpha_i$ to Eq. (45) is to correct the linear residual. Therefore, the procedure to determine $\alpha_i$ is called the linear residual correction (LRC).

Before presenting the LRC procedure, the notation of the correction factor is changed from $\alpha_i$ to $\alpha_i^{(1)}$ to distinguish from additional correction factor that will be introduced in Section 4.4 to recover the conservation. Similar to the positivity checking procedure, the LRC is also carried out within the sub-iteration. Therefore, the LRC procedure is to determine $\alpha_i^{(1),m}$ so that

$$\bar{U}_i^{n+1,m+1}\left(\alpha_i^{(1)}\right) = \bar{U}_i^n - \alpha_i^{(1),m} \tilde{R}_i^{n+1,m} \Delta t \in \bar{G}. \tag{58}$$

The LRC will be implemented for both the density and the pressure.

For density, it is computed by

$$\bar{\rho}_i^{n+1,m+1}\left(\alpha_{i,\rho}^{(1)}\right) = \bar{\rho}_i^n - \alpha_i^{(1),m} \tilde{R}_{i,\rho}^{n+1,m} \Delta t \tag{59}$$

If

$$\bar{\rho}_i^{n+1,m+1}(1) \in \bar{G},$$

we set

$$\alpha_{i,\rho}^{(1),m} = 1.$$

Otherwise, if

$$\bar{\rho}_i^{n+1,m+1}(1) \notin \bar{G},$$

then we require

$$\bar{\rho}_i^{n+1,m+1}\left(\tilde{\alpha}_{i,\rho}^{(1)}\right) = \bar{\rho}_i^n - \tilde{\alpha}_{i,\rho}^{(1),m} \tilde{R}_{i,\rho}^{n+1,m} \Delta t = \delta_\rho. \tag{60}$$

$\tilde{\alpha}_{i,\rho}^{(1)}$ is solved to be

$$\tilde{\alpha}_{i,\rho}^{(1),m} = \frac{\bar{\rho}_i^n - \delta_\rho}{\left|\tilde{R}_{i,\rho}^{n+1,m} \Delta t\right|}. \tag{61}$$

According to Eq. (3), it can be shown that if

$$\alpha_{i,\rho}^{(1),m} \leq \tilde{\alpha}_{i,\rho}^{(1),m},$$

the positivity of density can be guaranteed. In practical implementation, we choose

$$\alpha_{i,\rho}^{(1),m} = \tilde{\alpha}_{i,\rho}^{(1),m}. \tag{62}$$

Then we deal with the pressure. According to Eq. (58), the cell-average of the pressure is computed by

$$\bar{p}_i^{n+1,m+1}\left(\alpha_{i,p}^{(1),m}\right) = (\gamma-1)\left\{\begin{array}{l}\left(\overline{\rho E}_i^n - \alpha_{i,p}^{(1),m}\tilde{R}_{i,\rho E}^{n+1,m}\Delta t\right)\\ -\dfrac{1}{2\Omega}\iiint_\Omega\left[\begin{array}{l}\dfrac{\left(\overline{\rho u}_i^n - \alpha_{i,p}^{(1),m}\tilde{R}_{i,\rho u}^{n+1,m}\Delta t + \alpha_{i,p}^{(1),m}\beta_i^m(\delta\rho u)_i^{n+1,m}\right)^2}{\bar{\rho}_i^n - \alpha_{i,p}^{(1),m}\tilde{R}_{i,\rho}^{n+1,m}\Delta t + \alpha_{i,p}^{(1),m}\beta_i^m(\delta\rho)_i^{n+1,m}} +\\ \dfrac{\left(\overline{\rho v}_i^n - \alpha_{i,p}^{(1),m}\tilde{R}_{i,\rho v}^{n+1,m}\Delta t + \alpha_{i,p}^{(1),m}\beta_i^m(\delta\rho v)_i^{n+1,m}\right)^2}{\bar{\rho}_i^n - \alpha_{i,p}^{(1),m}\tilde{R}_{i,\rho}^{n+1,m}\Delta t + \alpha_{i,p}^{(1),m}\beta_i^m(\delta\rho)_i^{n+1,m}} +\\ \dfrac{\left(\overline{\rho w}_i^n - \alpha_{i,p}^{(1),m}\tilde{R}_{i,\rho w}^{n+1,m}\Delta t + \alpha_{i,p}^{(1),m}\beta_i^m(\delta\rho w)_i^{n+1,m}\right)^2}{\bar{\rho}_i^n - \alpha_{i,p}^{(1),m}\tilde{R}_{i,\rho}^{n+1,m}\Delta t + \alpha_{i,p}^{(1),m}\beta_i^m(\delta\rho)_i^{n+1,m}}\end{array}\right]dxdydz\end{array}\right\}$$

(63)

If

$$\bar{p}_i^{n+1,m+1}(1) \in \bar{G},$$

we set

$$\alpha_{i,p}^{(1),m} = 1.$$

Otherwise, if

$$\bar{p}_i^{n+1,m+1}(1) \notin \bar{G},$$

we use

$$\bar{p}_i^{n+1,m+1}\left(\tilde{\alpha}_{i,p}^{(1),m}\right) = \delta_p \qquad (64)$$

to solve for $\tilde{\alpha}_{i,p}^{(1),m}$. It is worth noting that Eq. (64) is a complex nonlinear equation without an explicit analytical solution. In the present paper, the Newton iterative method is used to solve Eq. (64). When the initial value of $\alpha_{i,p}^{(1),m}$ is 1, it can be proved that the Newton iterative method is convergent. Numerical tests indicate that the convergence speed of the Newton iterative method is very rapid, and the iteration well be convergent within 4 steps. Since the number of cells with

$$\bar{p}_i^{n+1,m+1}(1) \notin \bar{G}$$

is very small, the iterations do not lead to a considerable increase of the computational costs.

According to Eq. (4), since the pressure is a concave function of the conservative variables,

$$\alpha_{i,\rho}^{(1),m} \leq \tilde{\alpha}_{i,\rho}^{(1),m}$$

is sufficient for the positivity of pressure. In practical implementation, we choose

$$\alpha_{i,p}^{(1),m} = \tilde{\alpha}_{i,p}^{(1),m}. \tag{65}$$

Finally, $\alpha_i^{(1),m}$ is determined by

$$\alpha_i^{(1),m} = \min\left(\alpha_{i,\rho}^{(1),m}, \alpha_{i,p}^{(1),m}\right), \tag{66}$$

which ensures

$$\bar{U}_i^{n+1,m+1} \in \bar{G}.$$

By the convergence of the sub-iteration, $\alpha_i^{(1)}$ is determined so that

$$\bar{U}_i^{n+1} = \bar{U}_i^n - \alpha_i^{(1)} \tilde{R}_i^{n+1} \Delta t \in \bar{G} \tag{67}$$

according to Eq. (58). The LRC is very simple and efficient. Furthermore, it is proved in Section 4.6 that for a degree $k$ polynomial reconstruction

$$\alpha_i^{(1)} = 1 + O\left(\Delta t^2 + \Delta x^{k+1}\right). \tag{68}$$

Therefore, the LRC can realize the positivity-preserving solution without adversely affecting the order of accuracy.

## 4.4 The conservative linear residual correction

In this sub-section, the sub-iteration index $m$ will be omitted when no confusion is caused. One issue of the LRC is that for the cells and their direct neighbors, the application of the LRC will destroy the local conservation of the numerical schemes. Indeed, according to Eq. (42), $\alpha_i^{(1)} \tilde{R}_i^{n+1}$ in Eq. (67) is

$$\alpha_i^{(1)} \tilde{R}_i^{n+1} = \frac{1}{1+\phi}\left(\alpha_i^{(1)} R_i^{n+1} - \alpha_i^{(1)} \phi \frac{\bar{U}_i^n - \bar{U}_i^{n-1}}{\Delta t}\right), \tag{69}$$

where

$$\alpha_i^{(1)} R_i = \frac{1}{\bar{\Omega}_i} \sum_{f=1}^{N_f} \alpha_i^{(1)} F_{n,f} \tag{70}$$

according to Eq. (11). Considering that the interface $f$ is shared by cells $\Omega_i$ and $\Omega_j$, and usually $\alpha_i^{(1)} \neq \alpha_j^{(1)}$, we have

$$\alpha_i^{(1)} \boldsymbol{F}_{n,f} \neq \alpha_j^{(1)} \boldsymbol{F}_{n,f}$$

where $\alpha_j^{(1)} \boldsymbol{F}_{n,f}$ is corresponding to the modified residual of $\Omega_j$

$$\alpha_j^{(1)} \boldsymbol{R}_j = \frac{1}{\overline{\Omega}_j} \sum_{f=1}^{N_f} \alpha_j^{(1)} \boldsymbol{F}_{n,f}.$$

Therefore, the local conservation of the FV scheme is lost. When shock discontinuities are presented in the flow field, the loss of the conservation can sometimes lead to the numerical errors in the predicted shock strength and position. In this sub-section, a procedure will be introduced to recover the conservation.

To recover the conservation, we will determine another set of correction factor $\alpha_i$. Instead of applying $\alpha_i$ to the linear residual directly in terms of

$$\overline{\boldsymbol{U}}_i^{n+1} = \overline{\boldsymbol{U}}_i^n - \alpha_i \tilde{\boldsymbol{R}}_i^{n+1} \Delta t,$$

we use $\alpha_i$ to define the interfacial correction factor $\alpha_f$ using the relation

$$\hat{\alpha}_f = \min(\alpha_i, \alpha_j) \tag{71}$$

where the interface $f$ is shared by $\Omega_i$ and $\Omega_j$. And the corrected numerical scheme is changed to

$$\overline{\boldsymbol{U}}_i^{n+1} = \overline{\boldsymbol{U}}_i^n - \hat{\tilde{\boldsymbol{R}}}_i^{n+1} \Delta t \tag{72}$$

where

$$\hat{\tilde{\boldsymbol{R}}}_i^{n+1} = \frac{1}{1+\phi} \left( \hat{\boldsymbol{R}}_i^{n+1} - \alpha_i \phi \frac{\overline{\boldsymbol{U}}_i^n - \overline{\boldsymbol{U}}_i^{n-1}}{\Delta t} \right). \tag{73}$$

and

$$\hat{\boldsymbol{R}}_i^{n+1} = \frac{1}{\overline{\Omega}_i} \sum_{f=1}^{N_f} \hat{\alpha}_f \boldsymbol{F}_{n,f}. \tag{74}$$

And the positivity-preserving problem is to determine $\alpha_i$ so that $\overline{\boldsymbol{U}}_i^{n+1} = \overline{\boldsymbol{U}}_i^n - \hat{\tilde{\boldsymbol{R}}}_i^{n+1} \Delta t \in \overline{G}$.

This procedure is fully conservative, and is termed as the conservative linear residual correction (CLRC) procedure. For simplicity, the superscript $m$ corresponding to the sub-iteration is omitted in this sub-section.

The key issue of CLRC is how should $\alpha_i$ be determined. The most straightforward choice is to set

$$\alpha_i = \alpha_i^{(1)} \tag{75}$$

where $\alpha_i^{(1)}$ is determined by Eq. (66). If $\alpha_i$ is a local minimum within the compact stencil shown in Fig.1, i.e.

$$\alpha_i \leq \min\left(\alpha_{j_1}, \alpha_{j_2}, \alpha_{j_3}\right), \tag{76}$$

we have

$$\hat{\alpha}_f = \min\left(\alpha_i, \alpha_j\right) = \alpha_i = \alpha_i^{(1)} \tag{77}$$

since $\alpha_j$ is one of $\alpha_{j_1}, \alpha_{j_2}$ and $\alpha_{j_3}$. In this case,

$$\begin{aligned}
&\alpha_i \tilde{\boldsymbol{R}}_i^{n+1} - \hat{\tilde{\boldsymbol{R}}}_i^{n+1} \\
&= \frac{\Delta t}{\overline{\Omega}_i} \sum_{f=1}^{N_f} \left(\alpha_i - \hat{\alpha}_f\right) \boldsymbol{F}_{n,f} \\
&= \frac{\Delta t}{\overline{\Omega}_i} \sum_{f=1}^{N_f} \left(\alpha_i^{(1)} - \alpha_i^{(1)}\right) \boldsymbol{F}_{n,f} = \boldsymbol{0}.
\end{aligned} \tag{78}$$

Therefore, on cell $\Omega_i$, CLRC is equivalent to LRC, and the positivity can be guaranteed for numerical solution on cell $\Omega_i$. However, if Eq. (76) not valid, the choice of Eq. (75) is not positivity-preserving. A better choice is to ask that $\alpha_i$ is determined by

$$\alpha_i = \begin{cases} 1 & \text{if } \alpha_i^{(1)} = 1 \\ \min_{\forall \Omega_j}\left(\alpha_j^{(1)}\right) & \text{otherwise.} \end{cases} \tag{79}$$

We notice that Eq.(79) is also accuracy-preserving according to Eq.(68). We will prove that this choice of $\alpha_i$ can preserve the positivity of the density and pressure on almost all cells. To facilitate the discussion, the control volumes are classified into three groups. The first group is the set $N$ defined by

$$N = \{\Omega_i | \alpha_i < 1\},$$

the second group is the set $P$ satisfying

$$P = \{\Omega_i | \alpha_i = 1 \ \& \ \alpha_j = 1, \forall \Omega_j \in S_i \setminus \Omega_i\},$$

and the third group is the set $I$ characterized by

$$I = \{\Omega_i | \alpha_i = 1 \ \& \ \alpha_j < 1, \exists \Omega_j \in S_i \setminus \Omega_i\}.$$

It is clear that set $N$ consists of cells with possible negative density and pressure, set $P$ includes cells with positivity-preserving solutions, and set $I$ comprises the cells separating sets $N$ and $P$.

For $\Omega_i \in N$, we have

$$\alpha_i = \min_{\forall \Omega_j}\left(\alpha_j^{(1)}\right)$$

and

$$\hat{\alpha}_f = \min\left(\alpha_i, \alpha_j\right) = \alpha_i = \min_{\forall \Omega_j}\left(\alpha_j^{(1)}\right)$$

according to Eq. (79). Therefore,

$$
\begin{aligned}
&\alpha_i \tilde{\boldsymbol{R}}_i^{n+1} - \hat{\tilde{\boldsymbol{R}}}_i^{n+1} \\
&= \frac{\Delta t}{\overline{\Omega}_i} \sum_{f=1}^{N_f} \left(\alpha_i - \alpha_f\right) \boldsymbol{F}_{n,f} \\
&= \frac{\Delta t}{\overline{\Omega}_i} \sum_{f=1}^{N_f} \left(\min_{\forall \Omega_j}\left(\alpha_j^{(1)}\right) - \min_{\forall \Omega_j}\left(\alpha_j^{(1)}\right)\right) \boldsymbol{F}_{n,f} = \boldsymbol{0}.
\end{aligned}
\tag{80}
$$

This result shows, CLRC degenerates to LRC on $\Omega_i \in N$. Therefore, all cells in Set $N$ remains positivity-preserving.

For $\Omega_i \in P$, we have

$$\alpha_i = 1$$

and

$$\hat{\alpha}_f = \min\left(\alpha_i, \alpha_j\right) = 1.$$

In this case, no correction is needed since

$$\overline{\boldsymbol{U}}^{n+1} \in \overline{G}$$

for $\Omega_i \in P$.

For $\Omega_i \in I$, we have

$$\alpha_i = 1$$

and

$$\hat{\alpha}_f = \min(\alpha_i, \alpha_j) = \alpha_j.$$

Then we have

$$\tilde{R}_i^{n+1} - \hat{\tilde{R}}_i^{n+1} = \frac{\Delta t}{\overline{\Omega}_i}\sum_{f=1}^{N_f}(1-\hat{\alpha}_f)F_{n,f} = \frac{\Delta t}{\overline{\Omega}_i}\sum_{f=1}^{N_f}(1-\alpha_j)F_{n,f}. \tag{81}$$

According to the definition of the set $N$, there is at least one interface $f$ that the conservative correction factors on the face neighbored cells $\Omega_i$ and $\Omega_j$ are different, i.e., $\alpha_i = 1 \neq \alpha_j$. Therefore, we have

$$\tilde{R}_i^{n+1} \neq \hat{\tilde{R}}_i^{n+1}.$$

In this case, it is not straightforward to know if the solution on cell $\Omega_i$ is positivity-preserving. To achieve the positivity of the solution on the set $I$ cells, the following measurement is proposed. It should be noticed that the following discussions are only applied to the set $I$ cells.

Firstly, additional positivity check is performed in terms of Eq. (72). The algorithm of this positivity check is basically the same as that of Section 4.2. The only difference is the conservative linear residual $\hat{\tilde{R}}_i^{n+1}$ is used to estimate the density and pressure. The relevant equations are listed below.

The density estimation:

$$\overline{\rho}_i^{n+1,m+1,*} = \overline{\rho}_i^n - \hat{\tilde{R}}_{i,\rho}^{n+1,m}\Delta t. \tag{82}$$

The high-order accurate pressure estimation:

$$\overline{p}_i^{n+1,m+1,*} =$$

$$(\gamma-1)\left\{\left(\overline{\rho E}_i^n - \hat{\tilde{R}}_{i,\rho E}^{n+1,m}\Delta t\right) - \frac{1}{2\Omega}\iiint_\Omega \begin{bmatrix} \dfrac{\left(\overline{\rho u}_i^n - \hat{\tilde{R}}_{i,\rho u}^{n+1,m}\Delta t + \beta_i^m \delta(\rho u)_i^{n+1,m}\right)^2}{\overline{\rho}_i^n - \tilde{R}_{i,\rho}^{n+1,m}\Delta t + \beta_i^m \delta\rho_i^{n+1,m}} + \\ \dfrac{\left(\overline{\rho v}_i^n - \hat{\tilde{R}}_{i,\rho v}^{n+1,m}\Delta t + \beta_i^m \delta(\rho v)_i^{n+1,m}\right)^2}{\overline{\rho}_i^n - \hat{\tilde{R}}_{i,\rho}^{n+1,m}\Delta t + \beta_i^m \delta\rho_i^{n+1,m}} + \\ \dfrac{\left(\overline{\rho w}_i^n - \hat{\tilde{R}}_{i,\rho w}^{n+1,m}\Delta t + \beta_i^m \delta(\rho w)_i^{n+1,m}\right)^2}{\overline{\rho}_i^n - \hat{\tilde{R}}_{i,\rho}^{n+1,m}\Delta t + \beta_i^m \delta\rho_i^{n+1,m}} \end{bmatrix} dxdydz\right\}.$$

(83)

And the second-order accurate pressure estimation:

$$\overline{p}^{n+1,m+1,**} \approx (\gamma-1)\left[\overline{\rho E}^{n+1,m+1,*} - \frac{\left(\overline{\rho u}^{n+1,m+1,*}\right)^2 + \left(\overline{\rho v}^{n+1,m+1,*}\right)^2 + \left(\overline{\rho w}^{n+1,m+1,*}\right)^2}{2\overline{\rho}^{n+1}}\right]. \quad (84)$$

Secondly, a method to recover the positivity of the solution is presented. If $\overline{\rho}_i^{n+1,m+1,*} \in \overline{G}$ and $\overline{p}_i^{n+1,m+1,*} \in \overline{G}$, the corresponding solution on cell $\Omega_i$ is still positivity-preserving. If $\overline{U}_i^{n+1,m+1,*} \notin \overline{G}$, the solution of Eq. (72) is not positivity-preserving. In the first case, no action is required. In the second case, the simplest method to recover the positivity is to move the cell from set $I$ to set $N$. Since for all cells in set $N$, Eq. (80) guarantees that the CLRC will be the same as LRC.

In the practical implementation, we need only to introduce a new correction factor denoted as $\alpha_i^{(2)}$. This factor is determined by

$$\alpha_i^{(2)} = \begin{cases} 1 & \text{if } \overline{U}_i^{n+1,m+1,*} \in \overline{G} \\ 0.9999 & \text{otherwise.} \end{cases} \quad (85)$$

Then Eq. (79) is modified to

$$\alpha_i = \begin{cases} 1 & \text{if } \alpha_i^{(1)}=1 \text{ and } \alpha_i^{(2)}=1 \\ \min_{\forall \Omega_j}\left(\min\left(\alpha_j^{(1)},\alpha_j^{(2)}\right)\right) & \text{otherwise.} \end{cases} \quad (86)$$

When Eq. (86) is used to determine sets $P$, $N$ and $I$, it is clear $\Omega_i \in N$ if $\overline{U}_i^{n+1,m+1,*} \notin \overline{G}$.

Therefore, Eq.(85) is just used for moving the cells with $\bar{U}_i^{n+1,m+1,*} \notin \bar{G}$ from $I$ to $N$.

To avoid possible confusion, we notice that the CLRC procedure is applied together with the sub-iteration. In the actual implementation, in one sub-iteration, we will firstly use the $\alpha_i$ and sets $P$, $N$ and $I$ determined by the previous sub-iteration to compute the linear residual without correction based on Eq. (42), i.e.

$$\tilde{R}_i^{n+1} = \frac{1}{1+\phi}\left(R_i^{n+1} - \phi\frac{\bar{U}_i^n - \bar{U}_i^{n-1}}{\Delta t}\right)$$

for all cells, and the conservative linear residual based on Eq. (73), i.e.

$$\hat{\tilde{R}}_i^{n+1} = \frac{1}{1+\phi}\left(\hat{R}_i^{n+1} - \alpha_i \phi\frac{\bar{U}_i^n - \bar{U}_i^{n-1}}{\Delta t}\right)$$

for cells $\Omega_i \in N \bigcup I$. Secondly, we check the positivity in terms of $\tilde{R}_i^{n+1}$, and $\alpha_i^{(1)}$ is computed using the LRC procedure of Section 4.3. Thirdly, we check the positivity in terms of $\hat{\tilde{R}}_i^{n+1}$, and $\alpha_i^{(2)}$ is computed using Eq. (85). Finally, $\alpha_i$ is computed using Eq. (86) and the sets $P$, $N$ and $I$ are updated. No further corrections are performed in this sub-iteration. $\alpha_i$ and the updated sets are stored for the next sub-iteration.

Therefore, LRC is a step of CLRC. The other steps of CLRC required considerably less efforts than LRC. The central idea of CLRC is to expand the set $N$ by adding the cells in set $I$ that is not positivity-preserving in terms of the conservative linear residual. As the expansion of set $N$, the set $I$ cells that surrounds the set $N$ will move further away from a PLVM that is usually located at the center of set $N$. Therefore, the validity of the present positivity-preserving algorithm relies on the following conjecture.

**Conjecture**. *During the sub-iteration, as the expanding of set $N$, there exists a corresponding set $I$, in which the solutions on all cells are with positive density and pressure in terms of the conservative linear residual.*

In practical simulations, this conjecture is always valid. The set $N$ expands at most twice to form the set $I$ with all positivity-preserving cells. The reason is very straightforward: Eq. (81) shows that the difference $\tilde{R}_i^{n+1} - \hat{\tilde{R}}_i^{n+1}$ is proportional to $1-\alpha_j$. Since $\alpha_j$ is very close to unity according to Eqs. (68), (85) and (86), and the cells in set $I$ are positive preserving in terms of $\tilde{R}_i^{n+1}$, these cells will be positivity-preserving as long as the solutions on these cells can tolerate the flux disturbances in the order of $1-\alpha_j$. Usual problems are with only some isolated points where negative density and pressure may appear, and the density and pressure are considerably larger than zero elsewhere. In this case, this conjecture will be true surely. One exception is that the flow fields is near vacuum everywhere, in which this conjecture may be problematic. However, this is not the problem that the present algorithm is intended to apply. Therefore, if the negative pressure and density only appear locally, and the pressure and density are not too small elsewhere, the present positivity-preserving algorithm will always work.

## 4.5 The time increment correction for LU-SGS iteration

The CLRC procedure ensures the positivity of the solution when the sub-iteration is convergent. However, the predicted solution at next pseudo time step in the LU-SGS sub-iteration may have negative density or pressure. In such scenario, a procedure similar to the LRC can be applied to the results of LU-SGS sub-iteration to enforce positivity of the intermediate iteration results.

Before the LU-SGS iteration, the residual of the dual time-stepping scheme, Eq. (16), is evaluated. After the application of the CLRC, the residual is

$$\widehat{R}_i^{n+1,m} = (1+\phi)\left( \hat{\tilde{R}}_i^{n+1,m} + \frac{\overline{U}_i^{n+1,m} - \overline{U}_i^n}{\Delta t} \right). \tag{87}$$

For cells $\Omega_i \in N \cup I$, the residual of Eq. (87) is used in Eq. (15) to solve for $\Delta \overline{U}_i^{n+1,m}$.

Before updating the solutions using Eq. (17), we need to check the positivity of the solution

$$\overline{U}_i^{n+1,m+1,*} = \overline{U}_i^{n+1,m} + \Delta \overline{U}_i^m.$$

This positivity check is very similar to the algorithm presented in Section 4.2. The related formulations are given below.

$$\overline{\rho}_i^{n+1,m+1,*} = \overline{\rho}_i^n - \Delta \rho_i^{n+1,m} \Delta t,$$

and

$$\bar{p}^{n+1,m+1,*} \approx (\gamma-1)\left[\overline{\rho E}^{n+1,m+1,*} - \frac{\left(\overline{\rho u}^{n+1,m+1,*}\right)^2 + \left(\overline{\rho v}^{n+1,m+1,*}\right)^2 + \left(\overline{\rho w}^{n+1,m+1,*}\right)^2}{2\bar{\rho}^{n+1}}\right].$$

Since $\Delta \bar{U}_i^m \to 0$ upon the convergence of the iteration, the second-order estimation of the pressure $\bar{p}^{n+1,m+1,*}$ is sufficient.

If

$$\bar{U}_i^{n+1,m+1,*} = \bar{U}_i^{n+1,m} + \Delta \bar{U}_i^m \in \bar{G},$$

we set the updated conservative variables as

$$\bar{U}_i^{n+1,m+1} = \bar{U}_i^{n+1,m+1,*}.$$

Otherwise, the correction factor $\alpha'$ is applied so that

$$\bar{U}_i^{n+1,m+1} = \bar{U}_i^{n+1,m} + \alpha_i' \Delta \bar{U}_i^m. \tag{88}$$

The procedure to determine $0 < \alpha_i' \leq 1$ is similar to the LRC procedure of Section 4.3 with Eqs. (89) and (63) being replaced by

$$\bar{\rho}_i^{n+1,m+1}\left(\alpha_{i,\rho}'^m\right) = \bar{\rho}_i^{n+1,m} + \alpha_{i,\rho}'^m \Delta \bar{\rho}_i^m$$

and

$$p_i^{n+1,m+1}\left(\alpha_{i,p}'^m\right) = (\gamma-1)\left\{\overline{\rho E}_i^{n+1,m} + \alpha_{i,p}'^m \Delta \overline{\rho E}_i^{n+1,m} - \frac{1}{2}\left[\frac{\left(\overline{\rho u}_i^{n+1,m} + \alpha_{i,p}'^m \Delta \overline{\rho u}_i^{n+1,m}\right)^2}{\bar{\rho}_i^{n+1,m} + \alpha_{i,p}'^m \Delta \bar{\rho}_i^{n+1,m}} + \frac{\left(\overline{\rho v}_i^{n+1,m} + \alpha_{i,p}'^m \Delta \overline{\rho v}_i^{n+1,m}\right)^2}{\bar{\rho}_i^{n+1,m} + \alpha_{i,p}'^m \Delta \bar{\rho}_i^{n+1,m}} + \frac{\left(\overline{\rho w}_i^{n+1,m} + \alpha_{i,p}'^m \Delta \overline{\rho w}_i^{n+1,m}\right)^2}{\bar{\rho}_i^{n+1,m} + \alpha_{i,p}'^m \Delta \bar{\rho}_i^{n+1,m}}\right]\right\}$$

respectively.

Since the second-order pressure is sufficient in this case, the procedure to compute $\alpha_{i,p}'^m$ is simpler than LRC. $\alpha_{i,p}'^m$ can be determined analytically without the need of iteration. $\alpha'$ is computed by

$$\alpha' = \min\left(\alpha_{i,\rho}'^m, \alpha_{i,p}'^m\right),$$

and the solution is updated using Eq. (88).

The present positivity-preserving algorithm can be summarized as the following algorithm. To show clearly how this procedure is implemented, a full sub-iteration step will be included.

**Algorithm (The positivity-preserving algorithm)**

*1: The operation of the algorithm starting with (LC) means loop over all cells

*2: The operation of the algorithm starting with (LE/LF) means loop over all edges/faces

*3: The operation of the algorithm starting with (LI) means loop for sub-iteration

- (LC) m=0, set $\alpha_j^{m=0}=1, \bar{U}_j^{n+1,m=0}=\bar{U}_j^n, P=\Omega, N=\varnothing, I=\varnothing$

- (LI) Start a sub-iteration, m=1 to M

- (LC) Reconstruct the reconstruction polynomials

- (LC) Apply the shock capturing limiter

- (LC) Apply the positivity-preserving limiter

- (LE/LF) Compute the numerical flux $F_{n,f}^m$; for $\Omega_i \cap \Omega_j = f$, if $\Omega_i \in N$ or $\Omega_j \in N$, compute $\hat{\alpha}_f^m = \min(\alpha_i^m, \alpha_j^m)$, and $\hat{\alpha}_f^m F_{n,f}$

- (LC) Compute the un-modified residual $R_i^{n+1,m} = \dfrac{1}{\bar{\Omega}_i}\sum_{f=1}^{N_f} F_{n,f}$ and the corrected residual

$$\hat{R}_i^{n+1,m} = \frac{1}{\bar{\Omega}_i}\sum_{f=1}^{N_f} \hat{\alpha}_f^m F_{n,f}$$

- (LC) Compute the linear residual

$$\tilde{R}_i^{n+1} = \frac{1}{1+\phi}\left(R_i^{n+1} - \phi\frac{\bar{U}_i^n - \bar{U}_i^{n-1}}{\Delta t}\right)$$

if $\Omega_i \in N \cup I$, compute the conservative linear residual

$$\hat{\tilde{R}}_i^{n+1,m} = \frac{1}{1+\phi}\left(\hat{R}_i^{n+1,m} - \alpha_i^m \phi\frac{\bar{U}_i^n - \bar{U}_i^{n-1}}{\Delta t}\right)$$

- (LC) Determine if $\bar{U}_i^{n+1,m+1} \in \bar{G}$ in terms of $\tilde{R}_i^{n+1,m}$ using algorithm of Section 4.2; if $\bar{U}_i^{n+1,m+1} \in \bar{G}$, set $\alpha_i^{(1),m+1}=1$, otherwise, compute $\alpha_i^{(1),m+1}$ in terms of algorithm of Section 4.3

- (LC) Determine if $\bar{U}_i^{n+1,m+1} \in \bar{G}$ in terms of $\hat{\tilde{R}}_i^{n+1,m}$; if $\bar{U}_i^{n+1,m+1} \in \bar{G}$, set $\alpha_i^{(2),m+1}=1$,

- otherwise, set $\alpha_i^{(2),m+1} = 0.9999$
- (LC) Compute $\min_{\forall \Omega_j}\left(\min\left(\alpha_j^{(1)}, \alpha_j^{(2)}\right)\right)$
- (LC) Compute

$$\alpha_i = \begin{cases} 1 & \text{if } \alpha_i^{(1)} = 1 \text{ and } \alpha_i^{(2)} = 1 \\ \min_{\forall \Omega_j}\left(\min\left(\alpha_j^{(1)}, \alpha_j^{(2)}\right)\right) & \text{otherwise.} \end{cases}$$

- (LC) Update $N, I, P$ according to $\alpha_i$
- (LC) Compute the residual for the LU-SGS iteration
- (LC) Perform the LU-SGS iteration
- (LC) Determine if $\bar{U}_i^{n+1,m+1} \in \bar{G}$ in terms of $\Delta \bar{U}_i^m$. If $\bar{U}_i^{n+1,m+1} \notin \bar{G}$, compute $\alpha_i'$ in terms of algorithm of Section 4.5
- (LC) Update the conservative variable $\bar{U}_i^{n+1,m+1} = \bar{U}_i^{n+1,m} + \alpha_i' \Delta \bar{U}_i^m$
- (LC) Check the convergence of the sub-iteration
- If the sub-iteration is convergent, exit loop (LI)
- End loop (LI)

## 4.6 Accuracy analysis

In this sub-section, we will prove that the LRC procedure is accuracy preserving. Since the correction factor of the CLRC is derived from the LRC, the CLRC will not affect the accuracy of the numerical schemes. For simplicity, only the density correction is considered. The analysis of the pressure correction is more complicated, but the result is the same. The numerical scheme for the density on a given control volume is

$$\bar{\rho}^{n+1} = \bar{\rho}^n + \frac{\theta}{1+\theta}\left(\bar{\rho}^n - \bar{\rho}^{n-1}\right) - \frac{1}{1+\theta} R_\rho\left(\bar{\rho}^{n+1}\right)\Delta t \qquad (90)$$

where $R_\rho$ is the residual of the FV scheme.

We consider the local truncation error (LTE). The exact solution of density is denoted as $\rho_e$. According to the definition of LTE, when substituting $\rho_e$ into the present FV scheme, we have

$$\frac{(1+\theta)\left(\bar{\rho}_e^{n+1}-\bar{\rho}_e^n\right)}{\Delta t} = \frac{\theta\left(\bar{\rho}_e^n - \bar{\rho}_e^{n-1}\right)}{\Delta t} - R_\rho\left(\bar{\rho}_e^{n+1}\right) + O\left(\Delta t^2\right) + O\left(h^{k+1}\right) \tag{91}$$

for a temporally second-order and spatially $k+1$-th order scheme, where $h$ is the mesh size. If we further assume that $R_\rho$ is Lipschitz continuous, it can be proved that

$$\bar{\rho}^{n+1} = \bar{\rho}_e^{n+1} + O\left(\Delta t^3\right) + O\left(h^{k+1}\Delta t\right) \tag{92}$$

if we assume $\bar{\rho}^n = \bar{\rho}_e^n$ and $\bar{\rho}^{n-1} = \bar{\rho}_e^{n-1}$. Since the LRC will be applied, we have therefore $\bar{\rho}^{n+1} < 0$. However, Eq.(92) shows that $\bar{\rho}^{n+1}$ is very close to $\bar{\rho}_e^{n+1} > 0$.

By applying LRC, the numerical scheme for the corrected solution $\bar{\bar{\rho}}^{n+1}$ becomes

$$\frac{(1+\theta)\left(\bar{\bar{\rho}}^{n+1}-\bar{\rho}^n\right)}{\Delta t} - \frac{\alpha\theta\left(\bar{\rho}^n - \bar{\rho}^{n-1}\right)}{\Delta t} = -\alpha R\left(\bar{\bar{\rho}}^{n+1}\right) \tag{93}$$

in which $\alpha$ is determined so that the converged solution is $\bar{\bar{\rho}}^{n+1} = \delta_\rho$. If $\delta_\rho$ is chosen properly, the LRC will not affect the accuracy of the numerical scheme. Indeed, if

$$\delta_\rho \in \left(0, \bar{\rho}_e^{n+1} - c\bar{\rho}^{n+1}\right]$$

where $c$ is a $O(1)$ positive number, we have

$$\bar{\bar{\rho}}^{n+1} = \bar{\rho}_e^{n+1} + O\left(\Delta t^3\right) + O\left(h^{k+1}\Delta t\right) \tag{94}$$

and

$$\bar{\rho}^{n+1} = \bar{\bar{\rho}}^{n+1} + O\left(\Delta t^3\right) + O\left(h^{k+1}\Delta t\right). \tag{95}$$

The Lipschitz continuity leads to

$$\bar{R}\left(\bar{\rho}^{n+1}\right) = R\left(\bar{\rho}^{n+1}\right) + O\left(\Delta t^3\right) + O\left(h^{k+1}\Delta t\right). \tag{96}$$

According to Eq. (93),

$$\alpha = \frac{\dfrac{1+\theta}{\Delta t}\left(\bar{\bar{\rho}}^{n+1} - \bar{\rho}^n\right)}{\dfrac{\alpha\theta}{\Delta t}\left(\bar{\rho}^n - \bar{\rho}^{n-1}\right) - \alpha R\left(\bar{\bar{\rho}}^{n+1}\right)}. \tag{97}$$

Substituting Eqs. (95) and (96) into Eq. (97), we have

$$\alpha = \frac{\dfrac{1+\theta}{\Delta t}\left(\bar{\rho}^{n+1} - \bar{\rho}^n\right) + O\left(\Delta t^2\right) + O\left(h^{k+1}\right)}{\dfrac{\alpha\theta}{\Delta t}\left(\bar{\rho}^n - \bar{\rho}^{n-1}\right) - \alpha R\left(\bar{\rho}^{n+1}\right) + O\left(\Delta t^3\right) + O\left(h^{k+1}\Delta t\right)}. \tag{98}$$

On the other hand, Eq. (91) and (92) lead to

$$\frac{(1+\theta)(\bar{\rho}^{n+1}-\bar{\rho}^{n})}{\Delta t}=\frac{\theta(\bar{\rho}^{n}-\bar{\rho}^{n-1})}{\Delta t}-R_{\rho}(\bar{\rho}^{n+1})+O(\Delta t^{2})+O(h^{k+1}). \tag{99}$$

Eqs. (98) and (99) give

$$\alpha=1+O(\Delta t^{2})+O(h^{k+1}). \tag{100}$$

Using Eq. (100), and substituting the exact solution into Eq. (93), we have again Eq. (91), which proves that, after applying LRC, the scheme is still temporally second-order accurate and spatially *k+1*-th order accurate.

# 5 Numerical tests

In this section, the proposed positivity-preserving is tested by solving several benchmark cases containing strong discontinuities or expansions. The cubic variational-reconstruction is used in the FV scheme resulting in a fourth-order scheme. The WBAP limiter [46]-[47] is used with a troubled cell indicator to capture discontinuities. The physical time step is uniform and fixed in each case, while the pseudo time step is determined locally with a CFL number of 10.

**5.1 Sedov blast-wave**

The Sedov blast-wave problem[35] describes the propagation of a point-blast wave, and the initial condition is

$$(\rho,u,p)=\begin{cases}(1,0,4\times10^{-13}) & 0<x<2-0.5\Delta x,\ 2+0.5\Delta x<x<4 \\ (1,0,2.56\times10^{8}) & 0<x<0.5\end{cases} \tag{101}$$

The spatial grid size $\Delta x=5\times10^{-3}$. We use four different sizes of physical time steps to evaluate the convergence process of sub-iterations. Fig. 2 shows the number of sub-iterations required during the time evolution. In this figure, the simulations are carried out until $t=10^{-3}$ with four different time-step sizes, and corresponding numbers of physical time-steps for these simulations are also different. The reference results in Fig. 2 are from Ref. [35]. According to Fig.2, at $\Delta t=2\times10^{-6}$, the sub-iteration converges at about 20 steps. As the increase of the time step size, the required number of sub-iterations increase. When the time step is as large as $\Delta t=1\times10^{-5}$, the sub-iteration will not converge until 100 sub-iterations. These results shows that the present positivity-preserving algorithms do not introduce restriction on the physical time step size. However, the physical time step will adversely affect the convergence of the sub-iteration when it is too large.

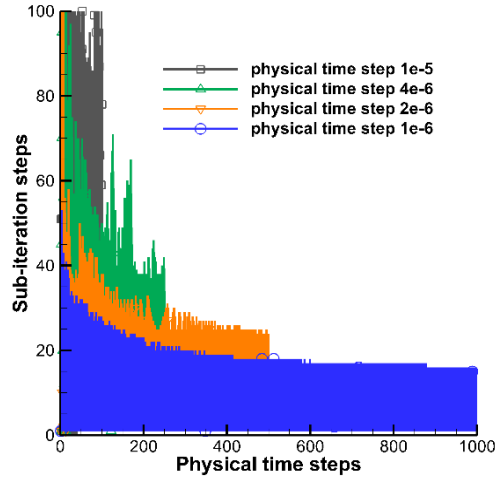

Fig. 2 The influences of the physical time step size on the convergence of the sub-iterations for the Sedov blast-wave problem.

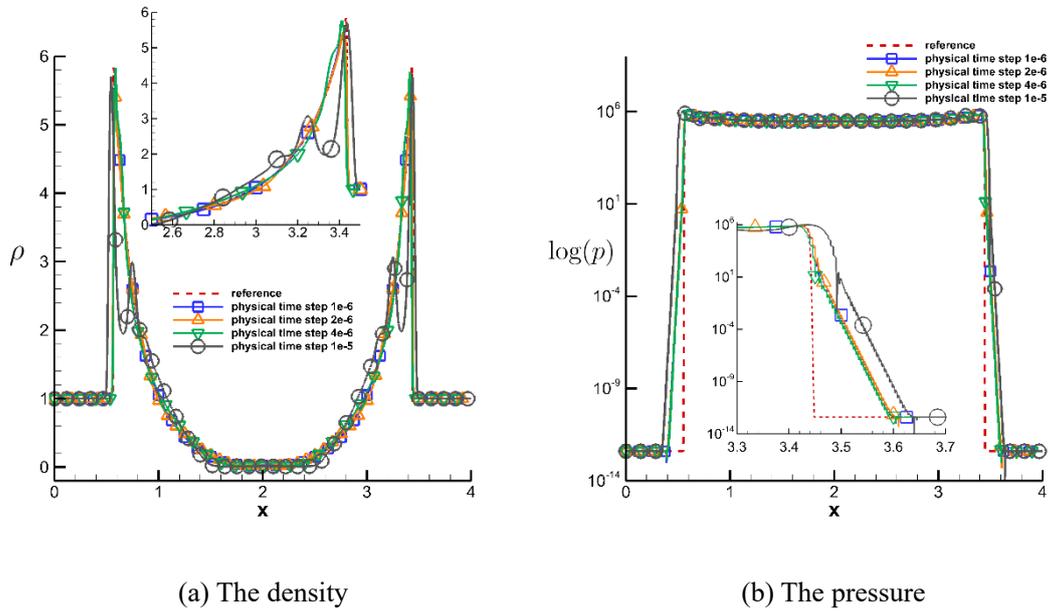

(a) The density  (b) The pressure

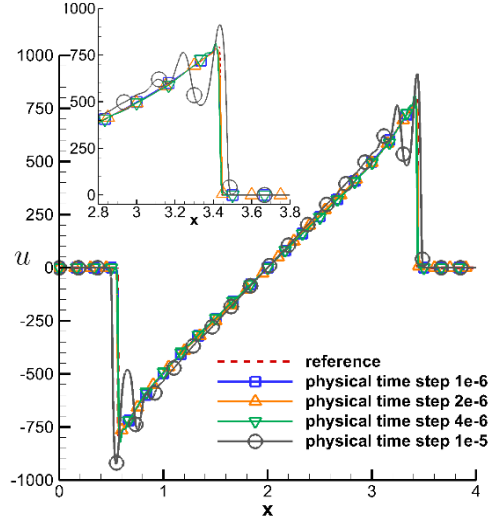

(c) The velocity

Fig. 3 The distributions of density and pressure of the Sedov blast-wave problem.

Fig. 3 presents the density, pressure and velocity distributions at different physical time step sizes. When the physical time step sizes are $\Delta t = 1\times 10^{-6}$ and $\Delta t = 2\times 10^{-6}$ corresponding to CFL numbers 5~10, the predicted solutions agree with the reference solutions well, When the time step increases to $4\times 10^{-6}$, there are distinguishable differences between the numerical results and the reference solutions. Small wiggles in the numerical solution can be observed. When the physical time step size reaches $\Delta t = 1\times 10^{-5}$ corresponding to the maximum CFD number being about 50, large fluctuations appear in the density and velocity distributions, and the errors in shock locations also increase. For all time steps, positivity-preserving solutions can be predicted showing the improvement in robustness after applying the present positivity-preserving algorithm.

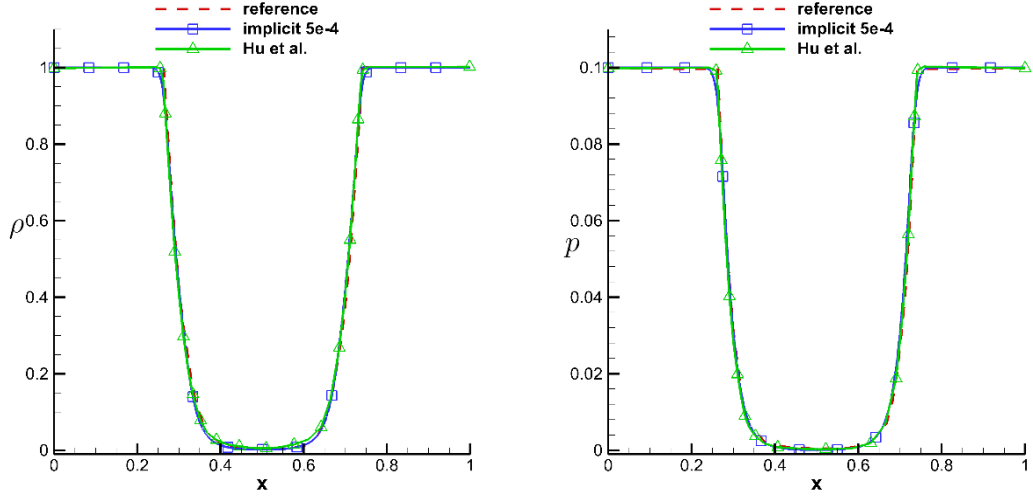

(a) The density            (b) The pressure

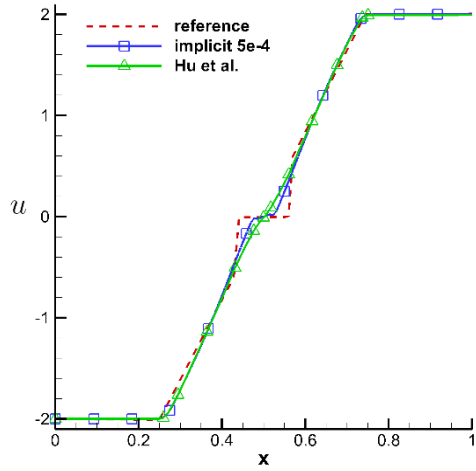

(c) The velocity

Fig. 4 The distribution of density, pressure and velocity of the double rarefaction problem

**5.2 Double rarefaction**

Double rarefaction problem [48] describes a one-dimensional shock tube problem involving a vacuum. The initial condition is

$$(\rho, u, p) = \begin{cases} (1, -2, 0.1) & 0 < x < 0.5 \\ (1, 2, 0.1) & 0 < x < 0.5. \end{cases} \quad (102)$$

The spatial grid size is $\Delta x = 2.5 \times 10^{-3}$, physical time step size is $\Delta t = 5 \times 10^{-4}$, and the simulation stops at $t = 0.1$. Fig. 4 presents the predicted density, pressure and velocity. The reference results in Fig. 4 are from Ref. [48], and the results of Hu et al. are from Ref. [38]. The predicted density and pressure agree with the reference solutions well. According to the distribution of velocity, we successfully capture the flat velocity distribution located at the near-vacuum region.

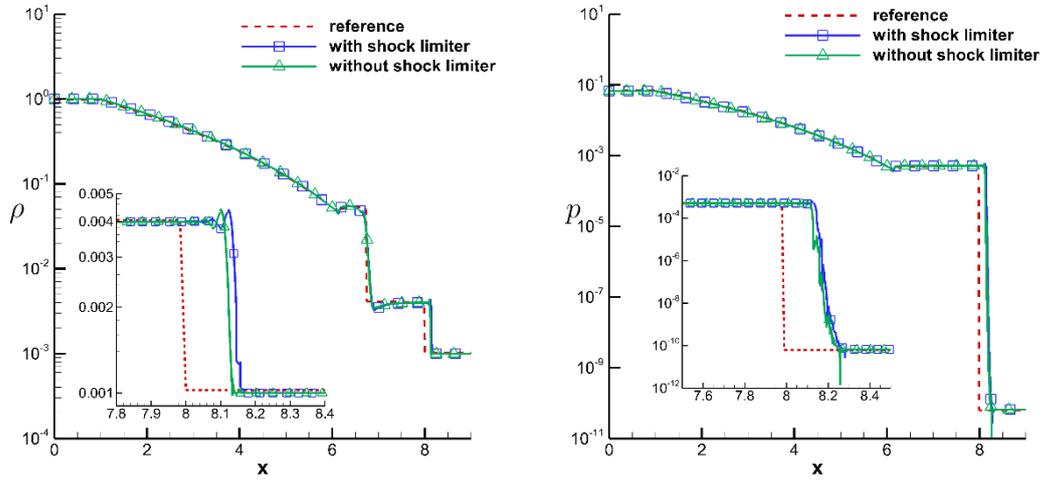

(a) The density	(b) The pressure

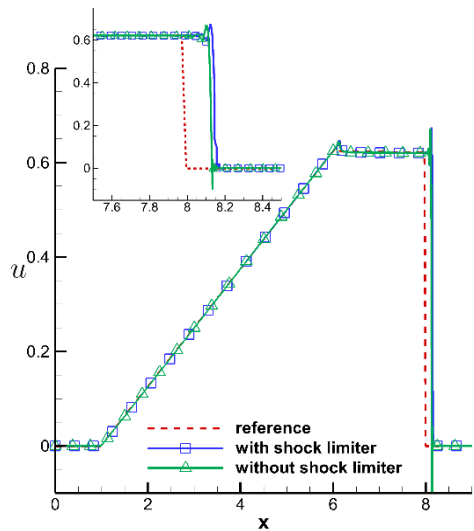

(c) The velocity

Fig. 5 The distribution of density, pressure and velocity of the Le Blanc shock-tube problem

**5.3 Le Blanc shock-tube**

Le Blanc problem [38] is a shock-tube problem with extremely small pressure. The initial condition is

$$(\rho, u, p) = \begin{cases} (1, 0, 2/3 \times 10^{-1}) & 0 < x < 3 \\ (10^{-3}, 0, 2/3 \times 10^{-10}) & 3 < x < 9 \end{cases} \quad (103)$$

The spatial grid size is $\Delta x = 9/800$, the physical time step size is $\Delta t = 6 \times 10^{-3}$, and solution is advanced in time till $t = 6$. In this case, we will study the influence of the limiter on the numerical solution when the present positivity-preserving algorithm is used. The numerical results are presented in Fig. 5. The reference results in Fig. 5 are from Ref. [38]. When the limiter is not applied, we can still obtain the positivity-preserving solution. However, noticeable oscillations in the vicinity of the discontinuities can be observed. When the limiter is turned on, the numerical oscillations can be controlled effectively.

**5.4 Mach 2000 jet**

The Mach 2000 jet problem [28] is a typical case to test the robustness of the scheme. In this paper, all two-dimensional test cases are in fact solved using a three-dimensional solver. Only one layer of control volumes is used in the z-direction. The computational domain is $[0,1] \times [-0.25, 0.25]$, and a uniform state $(\rho, u, v, w, p) = (0.5, 0, 0, 0, 0.4127)$ with a gas specific heat ratio $\gamma = 5/3$ is used to initialize the flow field. The upper, lower and right boundary conditions of the domain are outflow conditions, while the boundary condition on the left side of the domain is given by

$$(\rho, u, v, w, p) = \begin{cases} (5, 800, 0, 0, 0.4127) & y \in [-0.05, 0.05] \\ (0.5, 0, 0, 0, 0.4127) & \text{otherwise.} \end{cases} \quad (104)$$

The grid size is $\Delta x = \Delta y = \Delta z = 1/800$, and the physical time step size is $1 \times 10^{-6}$.

Fig. 6 presents the contours of density and pressure at $t = 0.001$ predicted by solving Euler equations and NS equations at Reynolds number $\text{Re} = 10^8$. It can be seen that the results of Euler equations and NS equations are similar in the compression region. However, in the expansion region, the results are quite different. We notice that the viscous flow results are reported for the first-time thanks to the proposed positivity-preserving algorithm that is applicable in a temporal implicit high-order NS solver.

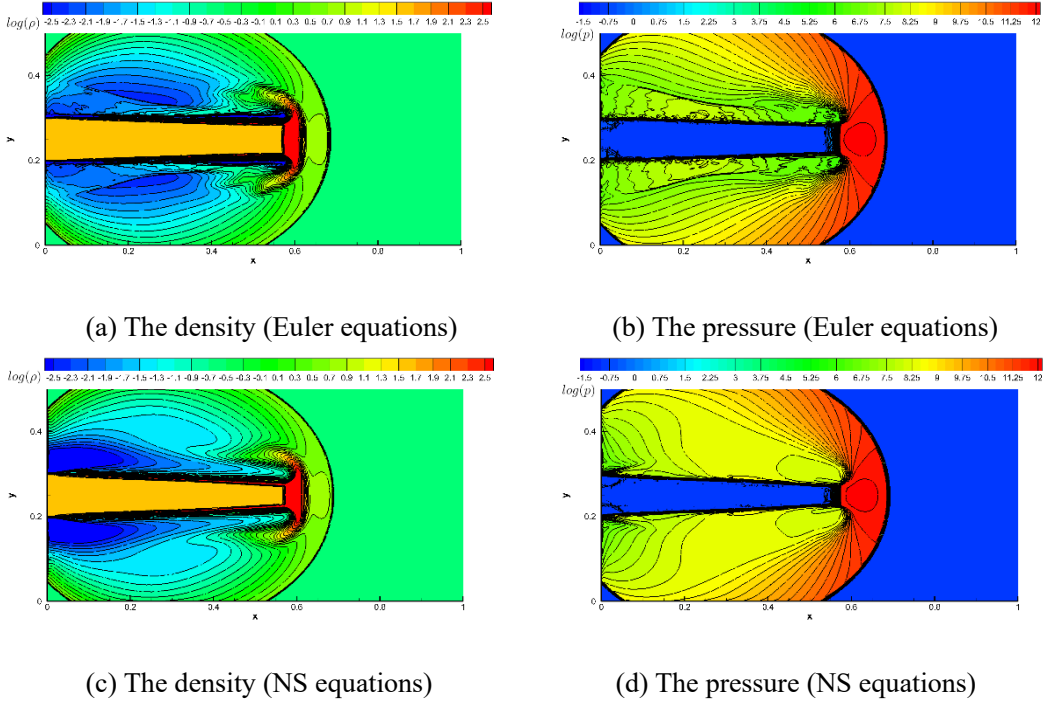

(a) The density (Euler equations)　　(b) The pressure (Euler equations)

(c) The density (NS equations)　　(d) The pressure (NS equations)

Fig. 6 Contours in logarithmic scale for the Mach 2000 jet problem. The viscous flow is solved at Reynolds number $\mathrm{Re}=10^8$.

**5.5 Shock reflection and diffraction**

Mach 10 shock reflection and diffraction problem [42] contains strong discontinuity and expansion. The initial condition is a right-moving Mach 10 shock wave located at x = 0.2. For $x > 0.2$, the flow field is a constant state with $(\rho,u,v,w,p) = (1.4,0,0,0,1)$. The upper and lower boundary conditions at $y = 0$ and $y = 2$ are the symmetrical conditions, and the wall boundary condition is applied at the surfaces of the wedge. The grid size $\Delta x = \Delta y = \Delta z = 1/120$, the physical time step size $1 \times 10^{-6}$, and the numerical results at $t = 0.001$ are presented.

Fig. 7 presents the predicted density and pressure obtained by solving the Euler and NS equations (Re = 100). Without using the positivity-preserving algorithm, the density and pressure will become negative near the sharp corner of wedge. By the introduction of the positivity-preserving algorithm, no negative density and pressure will appear. To further verify the positivity-preserving capability of the present method, we increase the shock Mach number to 100. The time step size is $\Delta t = 1 \times 10^{-7}$. Fig. 8 shows the numerical results at $t = 0.0001$. In this case, the current algorithm can still maintain the positivity of the density and pressure.

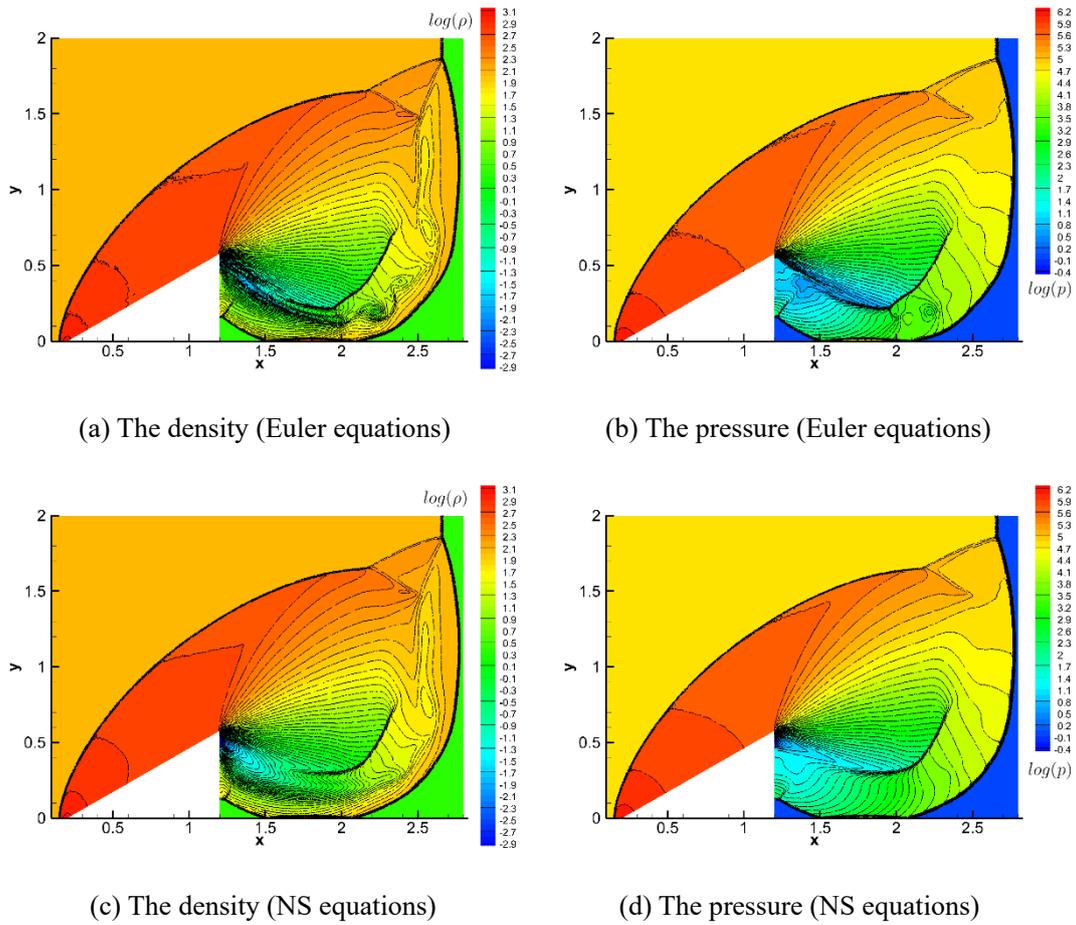

Fig. 7 The results in logarithmic scale for the Mach 10 shock reflection and diffraction problem.

Fig. 7 and Fig. 8 show that the inviscid and viscous flows have similar structures. However, the contact discontinuities predicted by the NS solver are weaker since the Reynolds number is rather low in this test case. When the shock Mach number increases, the structures of the shock and contact waves will become more complicated, which can however be captured by the fourth order accurate FV schemes.

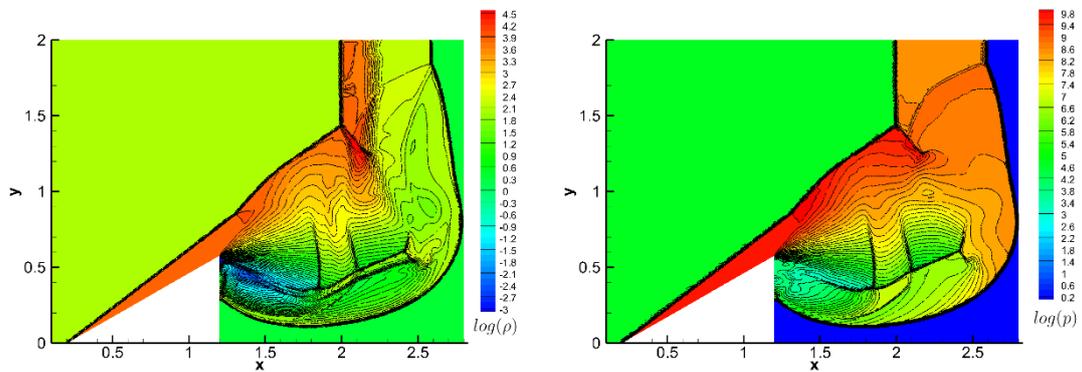

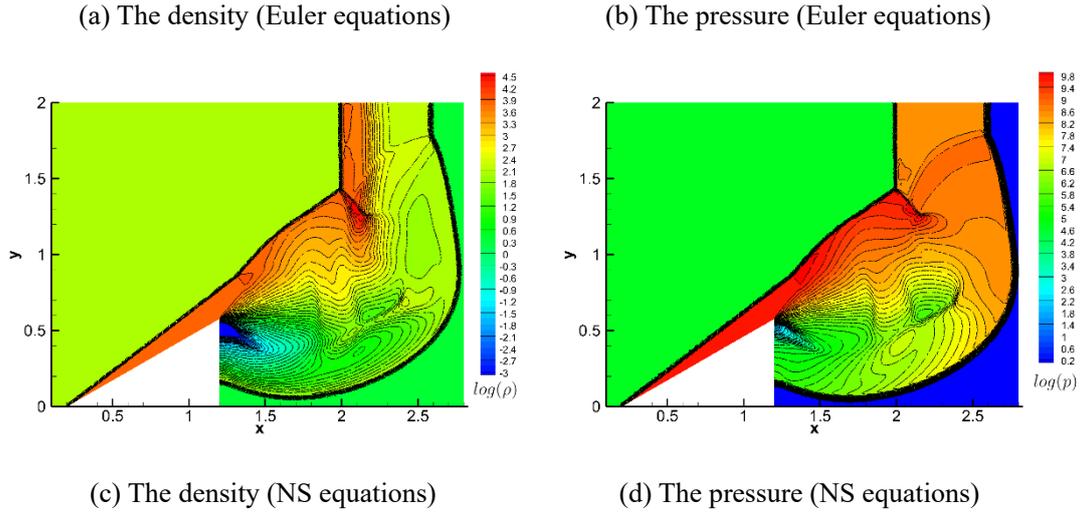

(a) The density (Euler equations)     (b) The pressure (Euler equations)

(c) The density (NS equations)     (d) The pressure (NS equations)

Fig. 8 The results in logarithmic scale for the Mach 100 shock reflection and diffraction problem.

# 6 Conclusions

Deterioration of robustness is one of the key issues restricting the application of high-order methods in engineering problems. For Euler and NS equations, the robustness issue is primarily embodied in that the positivity of density and pressure cannot be preserved during the computation, resulting in an abnormal termination of simulation. In this paper, we present a general positivity-preserving algorithm that is based on the physical consideration in terms of asymptotic analysis of the solutions of the Euler and NS equations near the local vacuum minimum points. The analysis shows that the solutions decay exponentially with time to maintain the non-negative density and pressure at a local vacuum minimum point. In its neighborhood, the exponential evolution leads to a modification of the linear evolution process, which can be modelled by a direct correction of the linear residuals of the numerical schemes to ensure positivity. A linear residual correction (LRC) procedure is then proposed. However, this procedure destroys the conservation of the numerical schemes. In the present paper, a conservative linear residual correction (CLRC) procedure is further proposed to recover conservation and maintain positivity of the solution simultaneously. The LRC is a step embedded in the CLRC procedure and requires the most amount of computation. The other parts of the CLRC demands considerably less computational efforts. Since the CLRC procedure is applied at only very small number of cells, the present positivity-preserving algorithm will not cause a considerable increase of computational cost. The proposed positivity-preserving algorithm is considerably less restrictive than existing algorithms. It does not rely on the existence of low-order positivity-preserving baseline schemes, convex decomposition of volume integrals of flow quantities or reduction of time step sizes. Furthermore, it can be implemented iteratively in the implicit dual time-stepping schemes to preserve positivity of the intermediate and converged states of sub-iterations. This algorithm can be also applied to explicit time-stepping schemes although it is not studied in this paper. Therefore, the present algorithm is one of the most general positivity-preserving methods till now. It is proved that the present positivity-preserving algorithm is accuracy preserving. Finally, the present algorithm has good scalability and can easily be extended to other numerical schemes besides the FV methods. The numerical tests demonstrate that the proposed algorithm predicts positive density and pressure, and maintains the convergence of the implicit temporal schemes.

# Acknowledgement

This work is supported by National Natural Science Foundation of China (12202246, 92152201) and the National Numerical Wind Tunnel Project.